\documentclass[twocolumn,showpacs,preprintnumbers,amsmath,amssymb,amsfonts]{revtex4}

\usepackage{graphicx}
\usepackage{dcolumn}
\usepackage{bm}
\usepackage{colordvi}
\usepackage{hyperref}

\begin{document}


\title{Quantitative imaging of concentrated suspensions under flow}

\author{Lucio Isa}
\author{Rut Besseling}%
\author{Andrew B Schofield}%
\author{Wilson C K Poon}%
\affiliation{Scottish Universities Physics Alliance (SUPA) and The
School of Physics and Astronomy, \\ The University of Edinburgh,
Kings Buildings, Mayfield Road, Edinburgh EH9 3JZ, United
Kingdom.\\}
\date{\today}

\begin{abstract}

We review recent advances in imaging the flow of concentrated
suspensions, focussing on the use of confocal microscopy to obtain
time-resolved information on the single-particle level in these
systems. After motivating the need for quantitative (confocal)
imaging in suspension rheology, we briefly describe the particles,
sample environments, microscopy tools and analysis algorithms
needed to perform this kind of experiments. The second part of the
review focusses on microscopic aspects of the flow of concentrated
'model' hard-sphere-like suspensions, and the relation to
non-linear rheological phenomena such as yielding, shear
localization, wall slip and shear-induced ordering. Both Brownian
and non-Brownian systems will be described. We show how
quantitative imaging can improve our understanding of the
connection between microscopic dynamics and bulk flow.

\end{abstract}

\pacs{83.80.Hj, 83.50.Ha, 83.60.La, 61.20.Ne, 61.43.Fs, 82.70.Dd}

\keywords{}

\maketitle

\tableofcontents

\narrowtext \noindent

\section{Introduction}

\begin{figure}
\includegraphics[width=0.3\textwidth,clip]{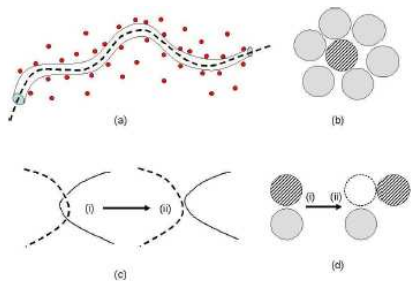}
\caption{A schematic comparison between polymer melt rheology and
colloid rheology. (a) In a polymer melt, a typical chain (dashed
curve) is constrained by many (in reality, $\sim 10^{3}$) other
chains, here represented by small circles. This gives rise to the
fruitful mean-field concept of a 'tube' in which the chain has to
move. (b) In a concentrated colloidal suspension, a typical
particle (hatched) is surrounded (in 3D) by $\sim 10$ neighbours.
This number is too small for mean-field averaging to be
meaningful. (c) Large deformations in polymer melts, such as the
process (i) $\rightarrow$ (ii), involves breaking covalent bonds,
and so do not ordinarily occur. (d) There are no covalent
constraints on order unity deformations, such as (i) $\rightarrow$
(ii), in a colloidal suspension.}\label{schematic}
\end{figure}

Understanding the deformation and flow, or rheology, of complex
fluids in terms of their constituents (colloids, polymers, or
surfactants) poses deep fundamental challenges, and has wide
applications \citep{LarsonBook}. Compared to the level of
understanding now available for the rheology of polymer melts, the
rheology of concentrated suspensions lags considerably behind. In
the case of polymer melts, it is now possible to predict with some
confidence flow fields in complex geometries starting from a
knowledge of molecular properties \citep{McLeishScience}. No
corresponding fundamental understanding is yet available for
concentrated colloids. The reasons for this are as follows
\citep{cates3}, Figure~\ref{schematic}.

In a polymer melt, each chain moves under the topological
constraints imposed by many other chains. The number of these
constraints, typically of order $10^3$, is sufficiently large that
a mean field picture, the so-called 'tube model', can be
successfully applied \citep{LarsonBook}. Moreover, the topological
entanglement between chains means that the breaking of covalent
bonds is needed to impose large deformations, so that strains
often remain small ($\ll 1$). In contrast, the maximum number of
neighbours in a (monodisperse) suspension of spheres is about 10,
so that 'mean-field' averaging of nearest-neighbour 'cages' will
not work, and local processes showing large spatio-temporal
heterogeneities are expected to be important. Moreover, no
topological constraints prevent the occurrence of strains of order
unity or higher, so that very large deformations are routinely
encountered. These two characteristics alone render suspension
rheology much more difficult. Added on top of these difficulties
is the fact that concentrated suspensions are in general
'non-ergodic', i.e. they are 'stuck' in some solid-like,
non-equilibrium amorphous state. On the other hand, in a
monodisperse system or in a mixture with carefully chosen size
ratios, highly-ordered (crystalline) states can occur. In either
case, any flow necessarily entails non-linearity (yielding, etc.).
Moreover, a suspension is a multiphase system, so that the
relative flow of particles and solvent can, and often does, become
important. This complication does not arise in polymer melts.
Finally, compared to the considerable effort devoted to the
synthesis and experimental study of very well characterized model
materials in polymer melts, corresponding work in colloid rheology
remains relatively rare. It is therefore not surprising that the
rheology of concentrated suspensions is not nearly as well
understood as that of polymer melts.

Advances in this field will pay rich dividends –- the successful
processing and application of concentrated colloids more often
than not depends on understanding, tuning and exploiting their
unique flow properties \citep{PasteReview}. But to build and
validate predictive theories of bulk rheological properties, we
need microstructural information. A 'case study' of this claim
comes from recent work on the qualitatively distinct rheologies of
so-called 'repulsive' and 'attractive' colloidal glasses.

Colloidal glasses are concentrated suspensions in which long-range
diffusion effectively vanishes \citep{sciortinoreview}. These
dynamically-arrested amorphous states have finite shear moduli in
the low-frequency limit. In slightly polydisperse hard-sphere like
suspensions, the transition from an 'ergodic fluid' (where
long-range diffusion is possible) to a glass occurs at a particle
volume fraction of $\phi \approx 0.58$
\citep{pusey1,MegenPRE98_tracersinglass}. The cause of dynamical
arrest is crowding. As $\phi$ increases, each particle spends
longer and longer being 'caged' by its nearest neighbours, until
at $\phi \approx 0.58$, the life time of these cages becomes
longer than any reasonable experimental time window. Each particle
can still undergo Brownian motion within its nearest-neighbour
cage, but its root mean-square displacement saturates at just over
$0.1a$ (where $a$ is the particle radius)
\citep{MegenPRE98_tracersinglass}, this quantity being a measure
of the time-averaged fluctuating 'cage size'. When oscillatory
shear strain is applied to a hard-sphere colloidal glass, yielding
occurs in a single-step process at a strain amplitude of just over
10\% \citep{PetekidisJPCM04_creep+flow}. This has been interpreted
in terms of 'cage breaking': the glass yields when
nearest-neighbour cages are strained beyond their 'natural'
(thermally induced) deformation.

\begin{figure}
\includegraphics[width=0.3\textwidth,clip]{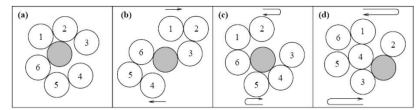}
\caption{Cartoon of two different yielding mechanisms in an
attractive glass under oscillatory shear (motion indicated by
arrows). (a) Un-sheared configuration of particles at time zero.
The shaded particle's bonding neighbours are particles 1, 2 and 6
and topological neighbours are particles 1–-6. At small shear
amplitude $\gamma_0$ the particles retain the same topology during
shear. (b) Above the lower yield strain, bonds break. (c) After
one period of oscillatory shear, the shaded particle retains the
same topological neighbours but different bonding neighbours, cf.
(a). (d) If the strain amplitude exceeds the higher yield strain,
the identity of the topological neighbours around the shaded
particle is changed after one period of shear, cf. (a), the glass
is melted and becomes liquid-like.}\label{fig:twostep}
\end{figure}

When a strong enough short-range attraction is present (where
'short' means a few per cent of $a$, a second type of glassy state
occurs –- the 'attractive glass', in which arrest is due to
particles being trapped by nearest neighbour 'bonds'
\citep{pham1}. It turns out that under oscillatory strain, an
attractive glass yields in a two-stepped process \citep{pham3}.
The first step occurs at a strain amplitude of a few per cent,
matching the range of the interparticle attraction, and then at a
second step when the strain amplitude reaches a few tens of per
cent. The first step has been interpreted as the breaking of
interparticle 'bonds', Figure~\ref{fig:twostep}. Once these are
broken many times, however, each particle 'realises' that it is
still in a topological cage formed by its neighbours. A second,
cage breaking', step is still necessary for complete yielding to
occur. This 'two step' signature is discernible in other
rheological tests applied to attractive colloidal glasses
\citep{PhamJRheol}.

For the purposes of this review, the important point is that the
explanation proposed for the observed qualitative difference
between repulsive and attractive glass rheology have been couched
in microstructural terms, Figure~\ref{fig:twostep}. Thus,
conclusive validation of these explanations depends on studying
yielding at the single-particle level. Such investigation can, in
principle, be performed using computer simulations. But simulating
particles under flow with realistic system sizes and including
hydrodynamic interactions is a major challenge, although progress
is being made (see the useful introduction in
\citep{PaddingLouis06} and references therein). Experimentally,
the situation is also difficult. Until comparatively recently, the
method of choice for obtaining microstructural information in
colloids is scattering (X ray, neutron or light), whether in
quiescent or flowing systems. But scattering can provides
information on the average structure; obtaining local information
is difficult, if not impossible. The ideal method for gaining
local information is direct imaging.

The appropriate imaging method of imaging colloids depends, of
course, on a number of factors, principally perhaps the size and
concentration of the particles concerned. The size imposes a basic
constraint, so that, for example, visible light can only be used
to image particles (at least in the `far field') with radius
$\gtrsim 300$~nm. The concentration is also an important factor,
because any multiple scattering by concentrated systems will
degrade image quality. In this review, we focus on the use of {\em
confocal} optical microscopy to study colloidal systems under
flow. Confocal imaging rejects out-of-focus information using
suitably placed pinholes, thus permitting the study of
concentrated systems. While the use of confocal microscopy to
study quiescent concentrated colloids is, by now, well established
\citep{dinsmore,habdas02,weeksrev1,weeksrev2}, the extension to
flow is relatively recent. The quantitative use of this
methodology depends on developments in both hardware and software.
First, to minimise image distortion, {\em fast} confocal scanning
is necessary; we review recent developments in this area in
Section~\ref{subsec:confocal}. Moreover, sample geometries are
required for simultaneous flow/rheometry and imaging, which we
review in Section~\ref{subsec:flowgeometries}. In terms of
software, algorithms must be developed to track particles from
image sequences distorted by flow; such algorithms are explained
in Section~\ref{sec:imageanalysis}. After discussing these
methodological issues, we turn to review a number of applications
(Section~\ref{sec:observations}).

To date, the use of quantitative confocal microscopy to image
concentrated colloids under flow has been applied mostly to the
study of very well characterised, model systems, particularly
systems of particles that interact as more or less perfect hard
spheres. The focus on model systems is partly dictated by the
desire to obtain as high quality images as possible, which
requires the matching of the refractive indices of the particles
and the suspending medium (see Section~\ref{subsec:particles}).
The degree of control necessary, both in the particle synthesis
and in solvent choice, is only really achievable in model systems.
The reason for choosing to study hard spheres is that these
represent the simplest possible (classical) strongly-interacting
particles. The properties of a quiescent collection of hard
spheres are well understood in statistical mechanical terms, much
of this understanding having been obtained in the last few decades
through the study of model hard-sphere colloids. It is therefore
natural to extend such studies into the area of {\em driven}
systems.

We first briefly summarize the quiescent behaviour of ideal hard
spheres. As first shown via numerical
simulations~\citep{wood,hoover}, below a volume fraction of
$\phi_{\rm F} = 0.494$, the thermodynamically stable phase is a
fluid of colloidal particles. Long-ranged order is absent and the
particles are able to explore all available space (the system is
ergodic). For $0.494 < \phi < 0.545$, the equilibrium state is a
coexistence of a fluid phase and a crystalline phase.  At
$\phi_{\rm C} = 0.545$ the entire system must be crystalline to
minimise the free energy, and this remains the case up to the
(crystalline) closed packing fraction of $\phi_{\rm CP} = 0.74$.
This behaviour was largely confirmed by the experiments of Pusey
and van Megen~\citep{pusey1} using sterically-stabilised PMMA
particles (for which see the next section). But, they also found
that their system failed to crystallize for volume fractions $\phi
\gtrsim \phi_{\rm G} = 0.58$, remaining `stuck' in a non-ergodic
(or glassy) state up to the maximum possible concentration for
amorphous packing (which, for monodisperse hard spheres, is
$\phi_{\rm RCP} \simeq 0.64$). We have already introduced such
colloidal glasses. Our review of microscopic phenomena will cover
both the flow of amorphous states (Section~\ref{subsec:disobserv})
and of ordered (or ordering) states (Section~\ref{subsec:order})
of concentrated colloids. While our focus will be on hard-sphere
colloids, we will also mention various imaging studies on the the
flow of non-Brownian suspensions.

\section{Experimental Methods}
\label{sec:Methods}

\subsection{The colloidal particles}
\label{subsec:particles}

One of the key ingredients for high-quality confocal imaging of
concentrated suspensions is a colloidal system that allows
identification of the individual spheres, does not suffer from
rapid photobleaching and gives good depth penetration into the
bulk of the dispersion. Fluorescently-labelled
poly-methyl-methacrylate (PMMA) particles in hydrocarbon solvents
show these characteristics.

The synthesis of hard-sphere PMMA particles was first described by
Barrett~\citep{barrett} in 1975 and subsequently by Antl
\textit{et al.}~\citep{antl} in 1986. It  is a two-stepped
dispersion polymerization reaction, yielding particles made of
PMMA cores kept stable by a thin ($\simeq$10nm) outer layer of
poly-12-hydroxystearic acid chains, which act as a steric barrier
to aggregation.  In the first step of the polymerization the
spheres are made by growing PMMA chains in solution, which become
insoluble when they reach a certain size.  At this point they come
out of solution and clump together to form particles that are kept
stable by physisorbed poly-12-hydroxystearic acid chains.  The
second step of the preparation involves chemically linking the
stabiliser chains to the spheres. These chains are tightly packed
on the particle surface~\citep{cebula} and stretch out in good
solvents such as various hydrocarbons, which causes the particles
to interact as nearly-perfect hard-spheres~\citep{bryant}. The
resulting particles can have polydispersities as low as a few
percent, and can be made with radii ranging from 100~nm up to and
above 1~$\mu$m.

Fluorescent labelling of the particles for confocal microscopy may
be achieved in three ways. The first involves the use of
polymerisable dyes. These dyes have been chemically modified to
include a reactive group that can be chemically attached to the
particle as they are produced. The advantage of this procedure is
that the dye will not leave the particle once it is incorporated.
For sterically-stabilised PMMA particles this involves adding a
methacrylate group to the dye, and several such
procedures~\citep{tse,liu2,tronc,jardine,bosma} have been
described in the literature. The most commonly used dye is
7-nitrobenzo-2-oxa-1,3-diazole-methyl methacrylate (NBD-MMA)
\citep{jardine,bosma}, which is excited at 488nm and emits at
525nm, while the red end of the spectrum is well served by
(rhodamine isothiocyanate)-aminostyrene (RAS) \citep{bosma}.

The second way to dye the PMMA spheres is to add an unreactive
species during particle formation. Here the dye has no
polymerisable group and is just dissolved in one of the reaction
reagents with the hope that it will become incorporated into the
growing particle. The advantage of this method is that no
chemistry, which may alter the dye's physical properties, is
required prior to use; it also allows for a wider range of dyes.
The disadvantage is that the dye may leak out when, for example,
solvency conditions are changed, or migrate within the particle.
This technique was employed by Campbell and
Bartlett~\citep{campbell}, who examined how four different red
dyes affected particle formation. They found optimum properties
when using 1,1-dioctadecyl-3,3,3,3-tetramethylindocarbocyanine
(DiIC18), which had significantly slower photobleaching rate than
other dyes tested, did not affect particle preparation and did not
interfere with the hard-sphere behaviour. However, this dye
degrades at temperatures around 100~$^{\circ}$C and therefore the
reaction that chemically links the stabiliser to the spheres can
not be performed~\citep{hu}.

The third method to stain the PMMA is to add the fluorescent dye
after particle synthesis. This is achieved by finding a solvent
that will dissolve the dye and also be taken up by the particles.
Thus, an acetone/cyclohexanone mixture can be used to deliver
rhodamine perchlorate dye to preformed PMMA
spheres~\citep{dinsmore}. The advantage of this method is that
once a suitable delivery system is found, many possible dyes, and
even multiple dyes, may be added to the spheres. The disadvantage
is that the solvent mixture may attack the spheres, swell them or
alter their physical properties.

An advantage of PMMA spheres is that the polymer may be
cross-linked \citep{pathmamanoharan1}. This is achieved via a
molecule with two polymerisable groups which is used to chemically
bind all the individual PMMA polymer chains in a particle together
into a network. This can stabilise the particles in solvents which
would normally dissolve them such as aromatics. An additional
benefit is that this method allows the fluorescent dye to be kept
within one particular area of the particle, usually its core. The
preparation of such core-shell particles
\citep{bosma,dullens1,dullens2,dullens3} involves modifying the
usual reaction by adding the cross-linking agent and dye at the
start of the procedure and creating particles as usual. However,
instead of going on to chemically link the stabiliser to the
particles, they are cleaned to remove excess dye and then more
methyl methacrylate monomer and cross-linking agent are reacted
with them to produce an undyed shell. The stabiliser is then
chemically attached. Such core-shell particles allow for more
accurate detection of the particle centers from microscopy in
concentrated systems, as the fluorescent cores are well separated
from each other (e.g. \citep{KegelScience2000}).

To achieve good imaging conditions in the bulk of the sample and
reduce scattering of the laser and excited light, the refractive
index (RI) of the particles and the solvent should be closely
matched. For PMMA spheres the RI is around 1.5 and can be matched
using a mixture of cis-decahydronaphthalene (RI=1.48) and
tetrahydronaphthalene (RI=1.54) in a ratio of approximately 2:1.
In earlier work~\citep{pusey1} a decalin-carbon disulphide mixture
was used (ratio 2.66:1) but this is problematic due to the
volatility and toxicity of the carbon disulphide. The particle RI
may also be modified by a few per cent by adding different
monomers during preparation~\citep{underwood}.

Another concern when studying colloidal dynamics is sedimentation.
To counter this phenomenon, solvent mixtures have been sought
which not only index-match the colloids but also allow density
matching. Adding carbon tetrachloride~\citep{dehoog} to the
cis-decahydronaphthalene/tetrahydronaphthalene system was tried
but had only limited success as it enhanced photobleaching and
imparted a charge to the particles. Another mixture used consists
of cis-decalin and cycloheptylbromide (CHB)~\citep{dehoog,weeks1}
(or cyclohexylbromide~\citep{leunissen}). This also imparts a
charge to the colloids, likely due to photo-induced cleavage of
the Br-C bond and subsequent solvent acidification, but this may
be screened by the addition of
salt~\citep{YethirajNature03_CHBmodelsystem,sedgwick1}, (partly)
restoring the nearly-hard-sphere nature of the system.

For both RI and density matching, it is important to note that the
extra solvent components can swell the PMMA particles, and the
solvent uptake may take up to weeks to saturate. Time-dependent
monitoring is the only sure means of ensuring that the particle
size has stabilised. Otherwise, significant errors in volume
fraction estimation may result.

Besides PMMA particles, various other systems have been considered
in the literature for use in confocal microscopy. One of the best
known is silica spheres. They are prepared by the hydrolysis and
condensation of alkyl silicates in ethanol using ammonia as
catalyst as first described by St\"{o}ber \textit{et
al.}~\citep{stober}. They can be made fluorescent by adding dyes
that have been reacted with silane coupling agents which make them
affix to the silica
\citep{vanblaaderen2,nyffenegger,verhaegh,imhof}. The advantage of
this system is that various dyes can be used and added at any time
during the synthesis \citep{vanblaaderen2}. The spheres are
stabilized in organic solvents via a dense layer of organophilic
material grafted onto their surface
\citep{vanhelden,pathmamanoharan2,pathmamanoharan3,vanblaaderen3},
and the RI can be matched with the solvent (RI=1.45 for St\"{o}ber
silica spheres). However, the main problem with silica spheres is
that they are quite dense (reported values range from
1.51g/cm$^3$~\citep{vanblaaderen3} to 2.2g/cm$^3$~\citep{vrij}) so
that sedimentation problems may be severe with all but the
smallest spheres.

Water-based fluorescent particles can also be used, but their RI
differs considerably from that of water, limiting observation in
the bulk. The RI of water can be modified by adding salts
\citep{hendriks}, but the concentrations required are
exceptionally high, making most particles unstable against van der
Waals attraction.

A final interesting development is that of quantum-dot-loaded
particles. Quantum dots \citep{murphy} are themselves very small
semiconducting colloids (1~nm to 10~nm) which fluoresce due to
quantum confinement. They can be trapped within a polymer or
silica~\citep{ma} colloid. Since quantum dots photobleach much
less than organic dyes, they can be used for experiments involving
long-term observation.

\subsection{Imaging}
\label{subsec:imaging}

Direct imaging dates back to the work of Robert Brown
\citep{brown}, who used it to discover and study Brownian motion,
which is {\em the} defining characteristic of colloids.
Subsequently, Perrin \citep{perrinBook} used direct imaging to
great effect in his Noble-Prize-winning work on sedimentation
equilibrium and diffusion in {\em dilute} suspensions.  However,
direct imaging has flourished as a major research tool in
colloidal research only in the last two or three decades, mainly
due to the increase in imaging and computing power.

Our main emphasis in this article is on high-resolution imaging
and reconstruction of the position of {\em every} particle in some
volume of a concentrated colloidal suspension under flow in
real-time. Before reviewing this subject in detail, we point out
the use of conventional, low-resolution imaging to track the
position of a number of {\em tracer} particles. Perhaps most
important for our purposes here, the well-established technique of
particle imaging velocimetry (PIV)~\citep{PIV} can be used to shed
light on complicating factors in conventional rheological
measurements such as wall
slip~\citep{RusselGrantColSurfA2000_slipyield,BuscalJRheo93_slip}
and flow non-uniformities such as shear banding \citep{Olmsted08}.
Thus this technique has recently been used in a rheometer to
elucidate the physics of wall slip in concentrated emulsions under
shear~\citep{MeekerPRL04_rheoslip,MeekerJRheo2004_slipandflowrheo}.
The correlation techniques used in PIV to measure tracer velocity
can also be used to track the shear-induced diffusion of tracers
in concentrated suspensions of non-Brownian
spheres~\citep{BreedveldJFlMech98,BreedveldPRE2001}. PIV and
related techniques are, of course, limited to transparent samples.

For completeness, we mention other methods for velocimetry that
have no requirement for transparency, such as heterodyne
light-scattering \citep{SalmonEPJAP2003_heterodyneDLS} and
ultrasonic velocimetry \citep{MannevilleEPJAP04_ultrasound}. The
latter has been applied to characterise slip and flow
non-linearities in micelles and
emulsions~\citep{BecuPRL06_twoemulsions,BecuPRE07_rheomic}. We
also mention under this heading Nuclear Magnetic Resonance Imaging
(NMRI)~\citep{fukushima_NMRAnnRevFlMech99,Callaghan_NMR_review_1999,BonnAnnRevFlMech2008_NMR,GladdenMeasSciTech96_NMR},
another velocimetry method independent of transparency, which can
also provide information on local density. The technique has
spatial resolution down to $\sim 20$~$\mu$m and has been combined
with rheometric set ups to relate velocity profiles to macroscopic
rheology~\citep{ovarlez2,RaynaudCoussotJRheo02_NMRyield}, to give
insight on the occurrence of shear
bands~\citep{HuangBonnPRL04_wetgranularflow} and shear
thickening~\citep{fall} in concentrated suspensions.

While all of these techniques give additional insight unavailable
from bulk rheology alone, building up a complete picture of
colloidal flow requires dynamical information on the
single-particle level. We now turn to microscopic methods that
give precisely such information. We focus on single-particle
imaging in 3D, but mention that the imaging of a single layer of
colloids has been used to great effect to study fundamental
processes in 2D (e.g.~\citep{Maret00,Maret03,Maret08}). While
perhaps somewhat less complex than 3D imaging, 2D imaging
nonetheless presents some challenges, e.g. when the imaged objects
come into very close proximity~\citep{Baumgartl}.

The use of conventional (non-confocal) optical microscopy to study
concentrated colloidal suspensions in 3D has been reviewed
before~\citep{Elliot01}. In nearly index-matched suspensions,
contrast is generated using either phase contrast or differential
interference contrast (DIC) techniques. One advantage of
conventional microscopy is speed: image frames can easily be
acquired at video rate. Conversely it has poor `optical
sectioning' due to the presence of significant out-of-focus
information, so that particle coordinates in concentrated systems
cannot be reconstructed in general, although structural
information is still obtainable under special
circumstances~\citep{Bristol97}.

Compared to conventional microscopy, confocal microscopy delivers
superior `optical sectioning' by using a pinhole in a plane
conjugate with the focal ($xy$) plane. It allows a crisp 3D image
to be built from a stack of 2D images even for somewhat turbid
samples, but each 2D image is acquired by scanning, which imposes
limits on acquisition speed. The technique has been described in
detail before~\citep{wilson}.

The use of confocal microscopy to study concentrated colloidal
suspensions was pioneered by van Blaaderen and Wiltzius
\citep{vanblaaderen1}, who showed that the structure of a
random-close-packed sediment could be reconstructed at the
single-particle level. Confocal microscopy of colloidal
suspensions in the absence of flow has been recently
reviewed~\citep{dinsmore,habdas02,weeksrev1,weeksrev2}. We refer
the reader to these reviews for details and references. Here, we
simply note that this methodology gives direct access to {\em
local} processes, such as crystal nucleation \citep{gasser} and
dynamic heterogeneities in hard-sphere suspensions near the glass
transition \citep{KegelScience2000,weeks1}.

Our main interest is the use of microscopy to study the flow of
concentrated suspensions at single-particle resolution in 3D. It
is possible to use conventional (non-confocal) video microscopy
for this
purpose~\citep{haw2,haw3,BiehlPalbergEPL04,SmithPetekidisPRE07_sheargel},
but the poor optical sectioning hinders complete, quantitative
image analysis. Conversely, crisp confocal images in principle
permit the extraction of particle coordinates, but due to slow
scanning and acquisition rates, early observations in real time
(i.e. during shear) produced blurred images that again limited the
potential for quantitative analysis~\citep{tolpekin}. A common
solution was to apply shear, and then image immediately after the
cessation of shear, both in 2D~\citep{hoekstra,stancik} and in
3D~\citep{VaradanSolomonJRheo2003_gelflowconfocal,tolpekin,CohenWeitzPRL04_confinedshear,SolomonJCP06_shearcrystalCF}.
In the next subsection we review developments in confocal
microscopy that permit faster acquisition and hence time-resolved
3D imaging of particulate systems under flow.

\subsubsection{Confocal microscopy}\label{subsec:confocal}

Confocal images are built by scanning a laser beam across the
field of view and collecting the emitted fluorescent light through
a pinhole (Laser Scanning Confocal Microscopy, LSCM).
Traditionally the laser beam is scanned across the specimen by two
galvo-mirrors which gives maximum acquistion rates of the order of
1 Hz, depending on image size. Technical advances such as the use
of resonant galvo-mirrors, spinning (Nipkow) discs (possibly
extended with an array of micro-lenses) and acousto-optic
deflectors (AODs) have significantly improved upon such
acquisition rates.

AODs are crystals which act as diffraction gratings. By sending a
standing sound wave at radio-frequency across the crystal the
local index of refraction is changed, creating a grating which
deflects laser beams passing through the crystal. By changing the
frequency of the sound wave, the diffraction angle is changed and
therefore the field of view can be scanned extremely rapidly. The
main problem associated with AODs is that the grating deflects
light of different wavelengths by different angles and therefore
obstructs the fluorescent light travelling back along the optical
path. This is partly resolved by combining an AOD with a
galvo-mirror and a slit instead of a pinhole. The galvo-mirror
positions the beam in one direction, the AOD scans in the
orthogonal direction and the fluorescent signal is detected
through the slit. The slit slightly reduces the rejection of
out-of-focus light and causes a slight anisotropy in the in-plane
point-spread function, but the method offers frame rates
$\gtrsim$~100~Hz for 512 $\times$ 512 pixels images.

Another technique that improves the scanning speed is the use of a
spinning (Nipkow) disc~\citep{wilson}. This solution operates by
illuminating the sample through an array of pinholes printed onto
a spinning disc and thus achieving the scanning of the sample
during the disc's revolution. Spinning discs can achieve full
scanning of the field of view by fractions of a revolution and
therefore yield frame rates of hundreds of Hz. Disadvantages of
this method are the fixed size of the pinholes, limiting the use
of different objectives for optimum operation, and also the strong
loss of intensity occurring at the disk. Recently, the last
disadvantage has mostly been overcome by using laser illumination
in combination with an additional array of micro-lenses, strongly
increasing the efficiency and considerably reducing
photo-bleaching \citep{WangDunn05}.

\begin{figure}
\includegraphics[width=0.3\textwidth,clip]{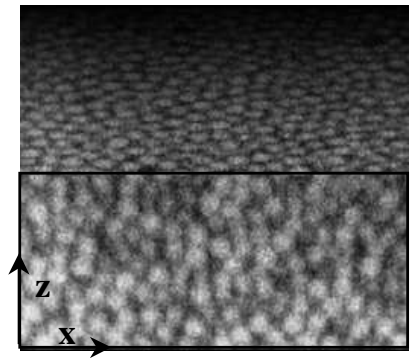}
\caption{Reprinted from~\citep{isa_methods}. Projection of a raw
3D image stack of 1.7~$\mu$m PMMA particles. The image volume, $x
\times y \times z \sim 29 \times 29 \times 15~\mu$m$^3$ ($256
\times 256 \times 76$ voxels), was scanned in $\sim 1$~s.}
\label{fig:3DCFimage}
\end{figure}

Regardless of the scanning principle, the confocal microscope can
be operated either in a 2D mode, i.e. capturing time sequences of
images at a fixed focal depth $z$, or in a 3D mode, i.e. capturing
image stacks obtained by scanning the sample volume,
Figure~\ref{fig:3DCFimage}. For rapid 3D acquisition, the best
method of scanning along $z$ is to use a piezo-element for focus
control of the objective. The two operation modes impose different
limits on the acquisition rates for successful quantitative
imaging (see Section~\ref{subsec:loc} and~\citep{isa_methods}).

\subsection{Flow geometries}
\label{subsec:flowgeometries}

Microscope studies of flowing particulate suspensions, and soft
matter systems in general, require well-defined flow geometries to
facilitate data interpretation and allow, in principle, for a
mapping of the observed dynamics to the rheological properties of
the system. To date, many studies have employed flow cells which
impose the deformation (rate) but do not have the ability to
measure directly the (shear) stress. Recently this limitation has
been overcome by combining confocal imaging with a rheometer and
by inferring (indirectly) stress from pressure drops in channel
flows.

In all geometries, wall properties play a crucial role in the
application of shear. The most direct manifestation of wall
effects is that of (apparent) slip in many complex fluids, in
particular colloid gels or pastes driven along smooth
surfaces~\citep{BallestaPRL08,MeekerPRL04_rheoslip,MeekerJRheo2004_slipandflowrheo}.
While the physics of slip in suspensions and pastes is an
interesting topic in its own right (see
Section~\ref{subsubsec:slowflow}), in most flow geometries the
practical goal is to minimise slip and transfer shear to the
suspension. The remedy for wall slip differs between various
systems; for moderate to very dense particle suspensions, a
coating consisting of a sintered monolayer of similar size
particles generally provides stick boundary conditions.

\subsubsection{Parallel plate shear cells}
\label{subsubsec:parallelplates}

Planar shear, or planar Couette flow, is simply implemented by
placing the material between two parallel plates much larger than
their separation $z_{gap}$, and translating them relative to each
other. This can be achieved either by fixing one of the plates and
moving the other one, or by moving them in opposite directions so
that, in the laboratory frame, the zero velocity plane is situated
somewhere between the two plates. For a Newtonian fluid without
slip and sufficiently far form the edges of the plates (a few
$z_{gap}$), the shear rate in the fluid is constant, $\dot
\gamma=(v_T-v_B)/z_{gap}$, with $v_T$ ($v_B$) the top (bottom)
plate velocity. In the following, we denote the shear velocity
direction by $x$, the vorticity direction by $y$ and the velocity
gradient direction by $z$.

By construction, parallel plate shear cells can only achieve
finite strains after which the direction of motion must be
reversed; they are thus particularly suited for oscillatory strain
studies. However, in some designs, extremely large strains are
possible ($\sim 20-50$) so that, with constant drive velocity,
steady shear is effectively achieved in each half-cycle. Using a
microscope glass (cover) slide as the bottom plate allows imaging
of the suspension during shear. An important requirement is that
the plates should be strictly parallel; this condition is
particularly stringent for small plate separations e.g.
\citep{WuImhof_RevSciInstr07}. Non-parallel plates induce
non-uniform shear as well as drifts in the sample due to
capillarity effects. Other considerations relate to component
weight, mechanical resonances, the driving motor and minimising
sample evaporation. Depending on the desired operation range,
compromises may be required to optimise either long-time stability
or high-frequency behaviour.

Planar shear cells were initially coupled to conventional
microscopes~\citep{haw2,haw3} or light scattering set
ups~\citep{Lequeux97,haw3,PetekidisFarDisc03_DWSrheo,PetekidisPRE02}.
With the advance of confocal imaging they are now preferentially
used in combination with an inverted microscope and a confocal
scanner.

Cohen {\it et
al.}~\citep{Cohen_oscill_shear_2006,Cohen_confined_crystal_2004}
used a simple design with a movable microscope cover slip as lower
plate and a fixed top plate. The maximum plate separation 100
$\mu$m and they were parallel to within $1\mu$m over the shear
zone. The lower plate was driven by a piezoelectric actuator with
displacements up to $90 \mu$m at frequencies $\leq$ 100 Hz. The
sample between the plates was in contact with a large reservoir of
un-sheared bulk suspension.

A similar design but allowing for larger strains and higher shear
rates is described by Solomon {\it et
al.}~\citep{SolomonJCP06_shearcrystalCF}. Two tilt goniometers
allow the user to tune the parallelism of the plates; a gap of
$150 \mu$m is set by a linear micrometer. Oscillatory shear was
produced by applying a sinusoidal displacement with a linear
stepper motor. The shear rates were in the range 0.01--100
s$^{-1}$ and strain amplitudes in the range 0.05--23.

A cell optimised for slow shear and large amplitudes was developed
by the Edinburgh group~\citep{BesselingPRL2007,isa_methods}. The
operational gap size ranged from $\sim 200-1000~\mu$m with plates
parallel to $\pm 5$~$\mu$m over the shear region. The top plate
was driven at $0.05-10~\mu$m$s^{-1}$ by a mechanical actuator with
magnetic encoder. The maximum translation was $L_s \sim 1$~cm,
allowing steady shear up to a total strain $\Delta
\gamma=L_s/Z_{gap} \gtrsim 1000 \%$. The cell could be operated
either with the bottom plate fixed or with the plates
counter-propagating via an adjustable lever system to tune the
zero velocity plane.

A recent, `state of the art', parallel plate cell is the design by
Wu {\it et al.}~\citep{WuImhof_RevSciInstr07}, which combines high
mechanical stability and a modular construction. The plate
separation ranges between 20 to 200$\mu$m, and a special pivot
system was designed to align the plates parallel to the highest
degree. A piezo-stepper motor provided plate velocities ranging
from $\sim 2~\mu$m$s^{-1}$ to $10$~mm$s^{-1}$. As in the
Edinburgh's design, the relative plate motion can reach up to
$\sim 1$ cm, allowing for large accumulated strain ($\lesssim 50$)
and quasi-steady shear.

\subsubsection{Rotational geometries}

\label{subsubsec:rotational}

Application of continuous shear is achieved in rotational
geometries such as cylindrical Couette, plate-plate or cone-plate
devices, the standard geometries used in traditional
rheology~\citep{Steffe}. Of these, cone-plate and plate-plate
geometries are most suitable for microscopic observation. Various
ways of conventional (microscopic) imaging in such geometries have
been used, including an early direct observation of
crystallization in a plate-plate rheometer
\citep{RodriguezKalerLangmuir93_crystalshear} and a microscopic
study of a confined, charged suspension under continuous shear in
a rotational plate-plate set-up \citep{BiehlPalbergEPL04}. Another
technique is to image tracer particles in the system from the side
of a cone-plate or plate-plate geometry to obtain the deformation
or velocity profile. This has been done both
qualitatively~\citep{AralJRheo94_surfaceroughness_slip} and
quantitatively~\citep{MeekerJRheo2004_slipandflowrheo}.

For completeness we also mention geometries used to study
non-Brownian suspensions, where the typical dimensions are much
larger (particle radius $a \gtrsim 50~\mu$m). One is an annular
channel formed by two concentric glass cylinders where a
ring-shaped top plate drives the suspension~\citep{TsaiPRL03}, the
other consists of a special truncated counter-rotating cone-plate
cell \citep{BreedveldPRE2001}. Both set ups allow for imaging in
multiple directions using different camera positions. As an
exponent of imaging with multiple cameras, we mention the recent
work in \cite{WangMaksePRE08}, where two orthogonally positioned
cameras were used simultaneously to image tracers in an immersed
granular packing sheared in a cylindrical Couette cell. Analysis
of the data from the two cameras provided the full
three-dimensional trajectory of the tracers.

\begin{figure}

\includegraphics[width=0.15\textwidth,clip]{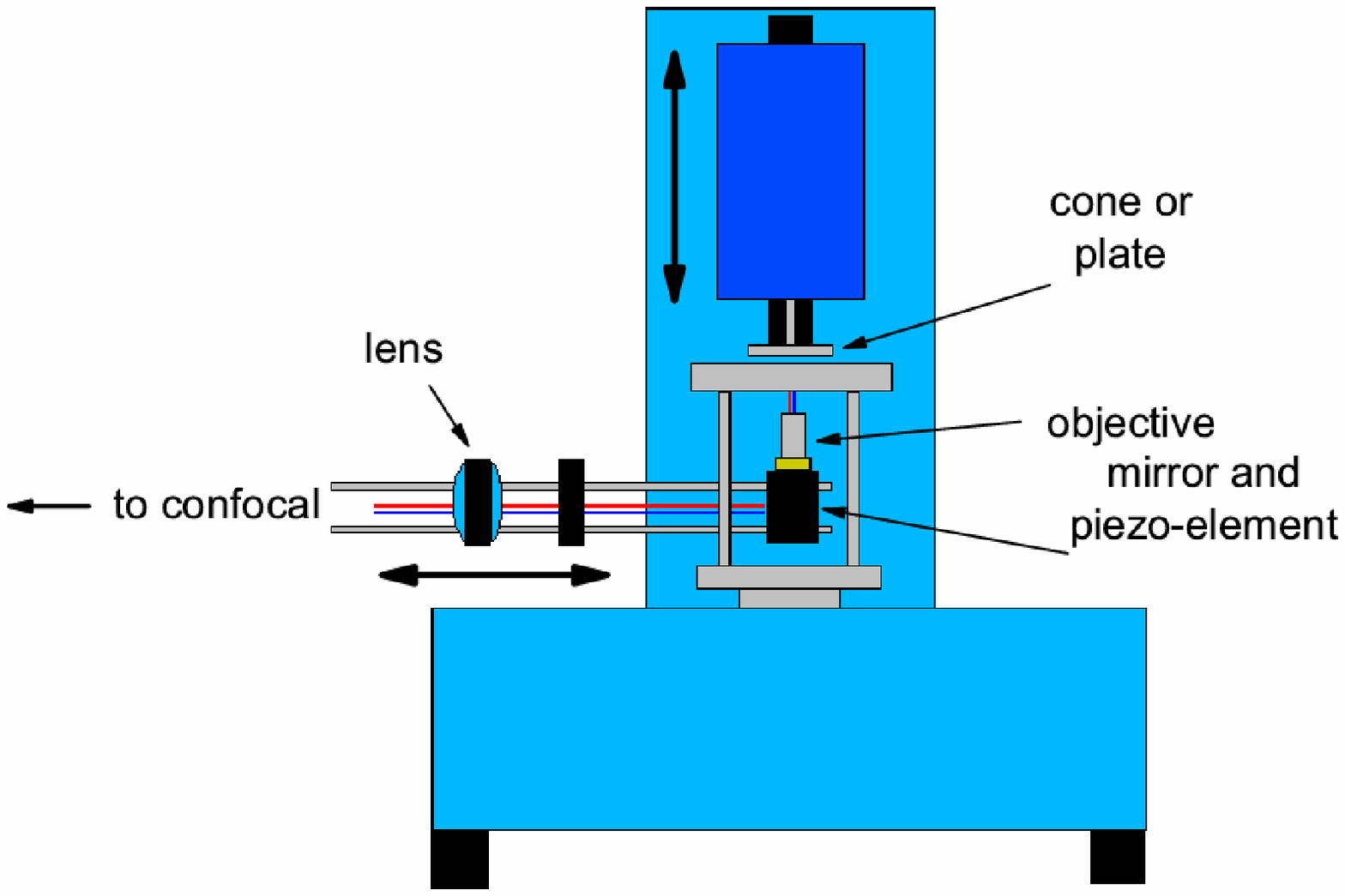}a)\includegraphics[width=0.15\textwidth,clip]{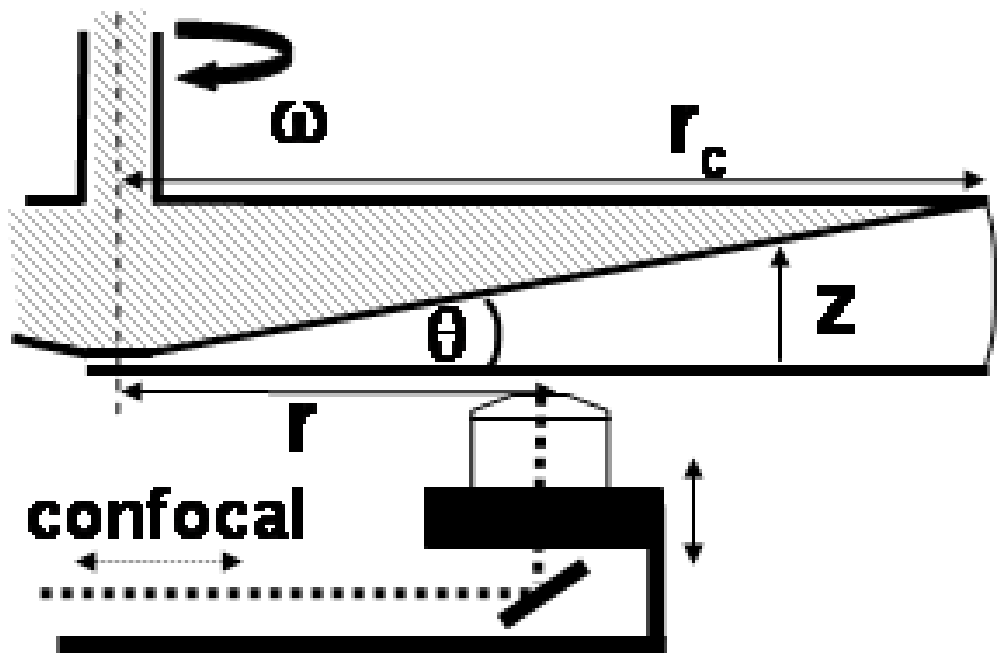}b)\includegraphics[width=0.15\textwidth,clip]{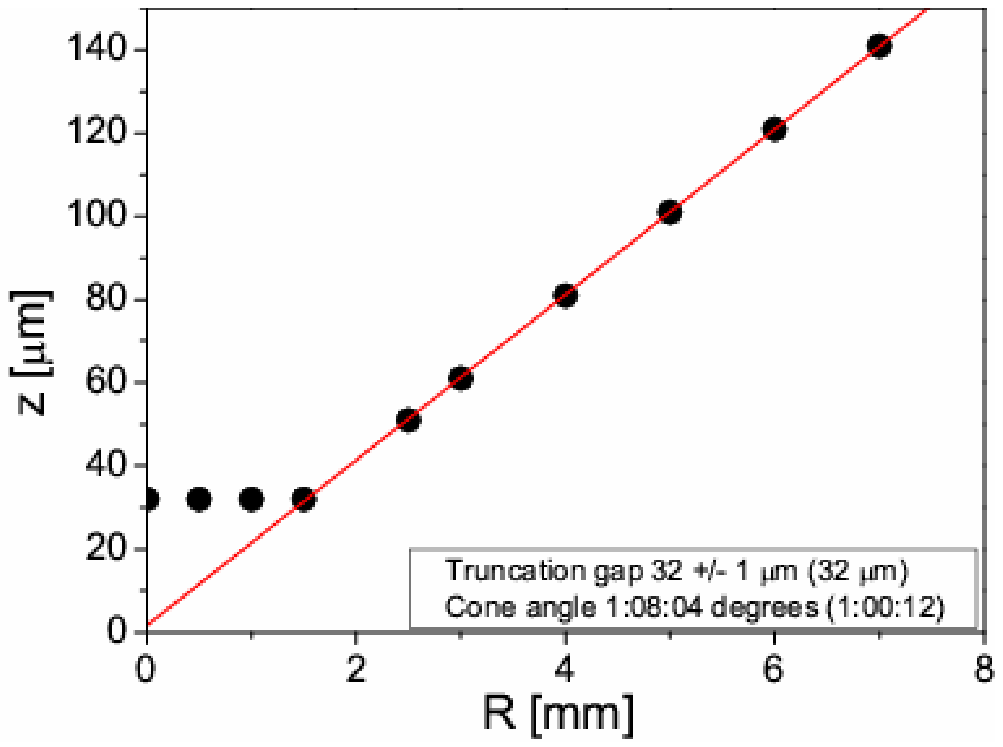}c)
\caption{Reprinted from~\citep{isa_methods}. (a) Schematics of the
confocal rheoscope of the Edinburgh group~\citep{isa_methods}. The
top arrow marks translation of the rheometer head to adjust the
geometry gap, the horizontal arrow indicates translation of the
arm supporting the objective to image at different radial
positions $r$. (b) Close up of the central part of the rheoscope,
similar to the cone-plate imaging system of
Derks~\citep{DerksJPCM04_coneplate} except that in the latter the
lower plate can also be rotated, while in the former the
microscope objective radial position $r$ can be varied. (c) Gap
profile of a $1^{\circ}$ cone-plate geometry, measured in the
confocal rheoscope with fluorescent particles coated on both
surfaces.} \label{fig:rheoscope}
\end{figure}

The first combination of a rotational shear cell with confocal
microscopy was described in
~\citep{nicolas,DerksJPCM04_coneplate}. It consists of a
cone-plate, similar to Figure~\ref{fig:rheoscope}(b), where the
cone and the plate (a large glass cover slide) could be rotated
independently. In contrast to Figure~\ref{fig:rheoscope}(b), the
objective was located at a fixed radial position where the gap
height was $1.7$~mm. This exceeds the working distance of high
numerical aperture objectives so that only the lower part of the
gap could be imaged at single-particle resolution, a general
limitation of any large-gap geometry. The shear rates obtained
were in the range of 10$^{-2}$ to 10$^2$ s$^{-1}$. The height of
the zero velocity plane could be adjusted by tuning the ratio of
the cone and plate rotational speed, while keeping the shear rate
constant. Due to the weight of the cone and plate, oscillatory
shear experiments were limited to low frequencies. The main
limitation of this set up is the acquisition rate of the confocal
scanner: particle tracking could be achieved only near the zero
velocity plane and in systems where the out-of-plane motion was
slower than the acquisition rate, such as s colloidal crystal.

A more recent development of confocal imaging in a `rheometric'
geometry, which also permits simultaneous measurement of the
rheological properties, is the set up of Besseling {\it et
al.}~\citep{isa_methods}, see Figure~\ref{fig:rheoscope}. Here a
rheometer was equipped with a custom-built, open base
construction, providing space for imaging optics. An adjustable
plate with an imaging slit at the top of the base is covered with
a circular cover slide forming the bottom surface of the geometry.
This allows for imaging at various radial distances $r$. The
stress-controlled rheometer (AR2000, TA instr.) enables all
classical rheological tests either in plate-plate or cone-plate
geometry. In practice a cone with an angle of 1$^\circ$ was used,
Figure~\ref{fig:rheoscope}(b),(c). The imaging optics under the
plate (piezo-mounted objective to scan along $z$, mirror and
lenses) are coupled to a confocal scanner (VTEye, Visitech) to
give full 3D imaging capabilities as in a standard microscope
mount. A major advantage is the acquisition rate (200 fps) of the
confocal scanner, which allows 3D imaging of single particle
dynamics up to rates of $\sim 0.1$~s$^{-1}$ and measurements of in
the $xy$ plane up to ~$100$~s$^{-1}$. Finally, either the cone or
the bottom glass plate (or both) can be coated with a sintered
layer of colloidal particles for the study of slip-related
problems.

\subsubsection{Capillaries and micro-channels}
\label{subsubsec:capillaries}

Micro-channels and capillaries are flow geometries that occur in
many practical applications. They are also interesting from a
fundamental perspective, offering insight on issues such as flow
instabilities ~\citep{HawPRL04_jamming},
confinement~\citep{isa_oscill} and particle migration
effects~\citep{abbott,leighton,frank,semwogerere1}. When the
pressure drop over the channel is measured, it is possible, in
principle, to relate microscopic observations to bulk rheological
properties (see e.g.~\citep{Degre_APL2006_microchannel}), but this
is non-trivial as it requires steady, uniform flow along the
channel and absence of entrance or confinement
effects~\citep{GoyonBocquet_nature2008_emulsion}.

From the 1970s, a large number of imaging studies of non-Brownian
suspensions flowing in mm- to cm-sized channels have been
performed via Laser Doppler
Velocimetry~\citep{koh,averbakh1,averbakh2,lyon} and Nuclear
Magnetic Resonance Imaging~\citep{sinton,hampton}, but very little
structural information has been obtained at the single-particle
level. Optical microscopy experiments on channel flows of colloids
have only recently started to appear, often in relation to
microfluidics applications~\citep{Microfluidics}. To avoid image
distortions, channels with square or rectangular cross sections
are generally preferred to cylindrical capillaries.

Haw~\citep{HawPRL04_jamming} studied the jamming of concentrated
hard-sphere colloids at the entrance of mm-sized, cylindrical
capillaries using conventional microscopy. The shape of the
channels and the method of imaging militated against accessing
detailed microscopic information.

Confocal studies in the group of Weeks~\citep{frank,semwogerere1}
used rectangular capillaries ($50 \mu$m$\times 500 \mu$m) coupled
to a syringe as a reservoir to drive suspensions of intermediate
volume fractions. Due to the high flow rates involved in their
experiments, particle tracking was impossible and instead the
flows were studied using image correlation techniques and
intensity measurements. A very similar geometry, but with square
channels was used in ~\citep{ConradLewisLangmuir08} to study the
flow of attractive gels of silica particles.

\begin{figure}
\includegraphics[width=0.3\textwidth,clip]{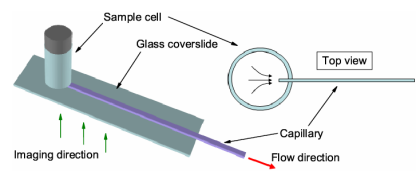}
\caption{A possible sample cell for capillary flow. The capillary
is not drawn to scale. The construction is placed on the
microscope stage and imaged from below. \label{fig:sample_cell}}
\end{figure}

Isa and co-workers~\citep{isa1,isa2,isa_methods} used a slightly
different geometry to study the flow of very dense colloidal
pastes. Figure~\ref{fig:sample_cell} shows a sketch of the setup:
square or rectangular, micron-sized glass capillaries are
connected to a glass reservoir at one end and flow is driven by
suction at the other end of the channel. The capillaries are
either untreated and smooth, or have the inner walls coated with a
sintered layer of PMMA colloids.

The recent development of micro-fabrication techniques (soft
lithography) has opened up the possibility of studying flow in
microfluidics geometries. Degr\`{e} {\it et
al.}~\citep{Degre_APL2006_microchannel} have performed PIV studies
of polymer solutions seeded with tracers in micro-fabricated
geometries made by adhering a moulded block of
polydimethylsiloxane (PDMS) onto a glass cover slide, obtaining
channels with a thickness of tens of microns. PDMS has many
practical advantages, but swells in most organic
solvents~\citep{lee2}, making it incompatible with many colloidal
model systems. It also has a relatively low elastic modulus
causing deformation under high pressures. A solution to these
problems is to substitute PDMS with a photo-curable monomer, as
described in ~\citep{bartolo}. It is clear that soft lithography
offers great flexibility in designing channels of almost any
geometry, paving the way for high-resolution imaging studies of
colloidal flows in a variety of complex micro-environments. Such
studies will be relevant for microfluidic applications, as well as
for modelling flows in porous and other complex materials.

\section{Image analysis}
\label{sec:imageanalysis}

To extract the maximum amount of quantitative information from
(confocal) images, detailed image analysis is necessary. The
central components of such analysis are particle location and
tracking.

\subsection{General methods}
\label{subsec:generalcorrel}

\begin{figure}
\includegraphics[width=0.3\textwidth,clip]{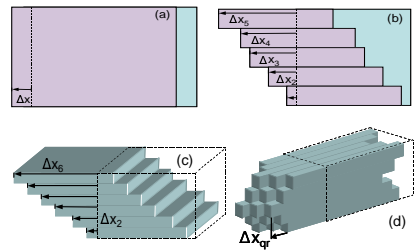}
\caption{Reprinted from~\citep{isa_methods}. Advection profiles in
various geometries. (a) A uniform 2D shift $\Delta X$ across the
entire field of view maximises the correlation. (b) 2D case where
the advected motion is a function of $y$; the image is then
decomposed in bins centered at $y_q$, each of which is shifted by
$\Delta X(y_q)$ to obtain maximum correlation. (C) 3D image where
the motion is a function of $z$ only. The stack is decomposed in
slices centered at $z_r$, each of which is shifted by $\Delta
X(z_r)$ to maximize the correlation. (d) 3D case with $y$ and $z$
dependent motion. Decomposition into $y$ and $z$ bins yields the
advection profile $\Delta X(y_q,z_r)$.}
\label{fig:advectionsketch}
\end{figure}

The basic method to obtain coarse-grained information on the flow
without details on the dynamics at the particle level is Particle
Image Velocimetry (PIV) or related correlation
techniques~\citep{PIV}. Using a sequence of images of tracer
particles in the suspension or confocal images of the full
microstructure during flow, a map of the advected motion between
consecutive 2D images, $i-1$ and $i$, or between parts of these
images (`tiles'), is obtained as the shift, $(\Delta X,~\Delta
Y)$, which maximizes the correlation between these images or
regions. By repeating the procedure over a sequence of frames, one
obtains the displacement as function of time, $(\Delta
X(t_i),\Delta Y(t_i))$. In general, PIV yields a discrete vector
field, $\mathbf{\Delta R}(\mathbf{r}_{pqr})$, with displacement
$\mathbf{\Delta R}$ in each element $\mathbf{r}_{pqr}$ of a 3D
image ($\mathbf{r}_{pq}$ in 2D). In many practical cases however,
the flow field has a simpler structure, see the examples in
Figure~\ref{fig:advectionsketch}.

For 2D images with uniform motion in the $xy$ plane, the procedure
is applied to the entire image,
Figure~\ref{fig:advectionsketch}(a), giving $\Delta { \bf R}({\bf
r},t_i)=\Delta X({\bf r},t_i)=\Delta X(t_i)$. For advective motion
which depends on the position $y$ transverse to the flow,
Figure~\ref{fig:advectionsketch}(b), the procedure is performed on
image strips, yielding $\Delta X(y_q)$ discretised at the strip
centers $y_q$. For 3D images of simple shear,
Figure~\ref{fig:advectionsketch}(c), the motion is a function of
$z$ only. The image stacks are then decomposed in $xy$ slices at
different $z$ and the procedure is applied to the individual 2D
slices. In channel flow, Figure~\ref{fig:advectionsketch}-d, the
stacks are first decomposed in $xy$ slices and then each slice is
decomposed in $y$-bins for which the motion is analysed.

These methods have been used to measure the velocity profiles in
various dense suspensions. Meeker {\it et
al.}~\citep{MeekerPRL04_rheoslip} imaged tracers embedded in
pastes from the edge of a cone-plate geometry. A number of studies
have been reported in which micro-PIV is applied to the flow in
channels~\citep{Degre_APL2006_microchannel,RobertsLewisLangmuir07_channelflow_gel}
or to obtain displacement profiles in parallel plate geometries
\citep{Cohen_oscill_shear_2006}.

\begin{figure}
\includegraphics[width=0.45\textwidth,clip]{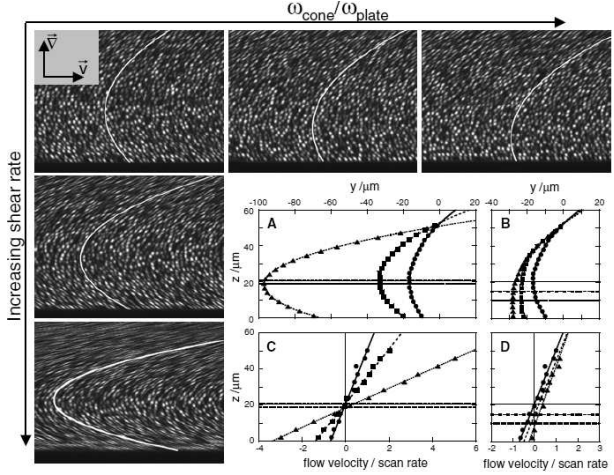}
\caption{Reprinted with permission from
~\citep{DerksJPCM04_coneplate}, copyright [2004], Institute of
Physics Publishing. Confocal images ($yz$,
75~$\mu$m$\times$56~$\mu$m, 512$\times$512 pixels$^2$) of a
colloidal fluid at various shear conditions taken in the
counter-rotating cone-plate shear cell
\citep{DerksJPCM04_coneplate}, see
Section~\ref{subsubsec:rotational}. The applied shear rates are
1.67, 3.36 and 8.39 s$^{-1}$ (top to bottom); the ratios of the
applied cone to plate rotation speeds are 84, 129 and 175 (left to
right). Graphs A and B show displacement profiles, $y(z)$,
measured from these images via cross-correlation of scanned lines.
The appropriate profile is overlaid on each image (white curves).
The velocity profiles ($dy/dz$) calculated from these displacement
profiles are shown in the graphs C and D. The particle diameter is
1.50$\mu$m. \label{fig:derks}}
\end{figure}

A related application of the correlation method has been reported
by Derks and co-workers~\citep{DerksJPCM04_coneplate}. In their
study of dense colloids flowing in a cone-plate geometry, a single
confocal scan in the velocity-gradient plane ($y-z$ plane in their
notation, see Figure~\ref{fig:derks}) was performed. This was then
analysed by shifting and correlating {\it lines} (along $y$)
between consecutive $z$-values in the image, rather than
performing PIV between consecutive frames. Via this procedure the
distortion of the particle images could be quantified, yielding
parabolic displacement profiles, Figure~\ref{fig:derks},
corresponding to linear velocity profiles from which the shear
rate was extracted.

\subsection{Locating particles}
\label{subsec:loc}

To go further in the analysis, one first needs to determine the
location of the particle centers from the images. The standard
method was developed over a decade ago by Crocker and Grier
(CG)~\citep{CrockerGrierJColIntSc96_tracking}, and has since been
used in numerous studies on colloidal dynamics.

While some of the concepts in their method may also apply to
identifying objects with varying shapes
\citep{BrangwynneKoenderinkBioPhysJ2007_filamenttracking}, the
method is primarily aimed at identifying circular (2D) or
spherical (3D) objects that appear bright on a dark background.
Since experimental images have unavoidable pixel noise and often
also undesired intensity modulations, the images are first treated
with a spatial bandpass filter, which eliminates long-wavelength
contrast gradients and pixel-to-pixel noise. Next, the coordinates
of the centers of the features are obtained by locating the local
intensity maxima in the filtered images. These coordinates are
then refined to a higher accuracy by applying a
\textit{centroiding} algorithm, which locates the
brightness-weighted center of  mass (centroid) of the particles;
with this refinement the coordinates of the particle center can be
obtained with a typical resolution of $1/n$ of the pixel
size~\citep{CrockerGrierJColIntSc96_tracking} where $n$ is the
particle diameter in pixels. However, the method  has some
limitations in concentrated systems, as individual particle images
may start to overlap. Unless one uses the core-shell particles
mentioned in Section~\ref{subsec:particles}, an alternative method
to the coordinate refinement may be required. A very useful
technique for doing this, based on optimizing the overlap between
the measured intensity profile of each particle and the so called
`sphere spread function', has been described recently by Jenkins
and Egelhaaf~\citep{JenkinsEgelhaafAdvColIntSci08}.

In addition to the intrinsic sub-pixel accuracy from the above
refinement methods, additional errors arise also from particle
motion during image acquisition. These errors may be considerable
when images are obtained by a scanning system, where the pixels
are not acquired instantaneously. For an acquisition time
$1/f_{\rm scan}$ for a 2D image with $n$ lines and a particle
radius (in pixels) of $\tilde{a}$, the acquisition time of a
particle image is:
\begin{equation}
\displaystyle{t_{\rm im}^{2D}=2\tilde{a}/(nf_{\rm scan})}.
\label{eq:ti2D}
\end{equation}
For a 3D image (a $z$-stack of 2D slices), the voxel size in the
$z$ direction may differ from that in the $x$ and $y$ direction.
When the particle radius in $z$-pixels is given by $\tilde{a}_{\rm
z}$, the acquisition time for a 3D particle image is:
\begin{equation}
\displaystyle{t_{\rm im}^{3D} = 2\tilde{a}_{\rm z}/f_{\rm scan}}. \label{eq:ti3D}
\end{equation}
Typical parameters ($f_{\rm scan}=90$~Hz in a fast confocal, $n=256$, $\tilde{a} = \tilde{a}_{\rm z}=5$) yield $t_{\rm im}^{2D} \simeq 0.4$~ms and $t_{\rm im}^{3D} \simeq 0.1$~s.

The additional errors resulting from this finite acquisition time
are easily estimated.  We do so here for hard spheres (HS). The
short time diffusion leads to an error $\delta_{HS} = \sqrt{D_{\rm
s}(\phi) t_{\rm im}/2}$. Here $D_{\rm s}(\phi)$ is the volume
fraction ($\phi$) dependent short time diffusion constant for hard
spheres: $D_{\rm s}(\phi)=(k_{\rm B} T)/(6 \pi \eta a) H(\phi)$
with $H(\phi)<1$ a hydrodynamic
correction~\citep{BeenhakkerMazurPhys84,pusey_hydro,TokuyamaPRE94_diffusion,BradyJChemPhys93_rheocol,MegenPRE98_tracersinglass}.
Note that this is an upper bound applicable to HS; for colloids
with softer interactions, $\delta < \delta_{HS}$. With a solvent
viscosity $\eta \simeq 3 \times 10^{-3}$~Pa$\cdot$s, and frame
rates as above, typical values are $\delta_{2D} \simeq 2$~nm and
$\delta_{3D} \simeq 35$~nm for a colloid with radius $a = 1~\mu$m.
For 3D imaging, this error exceeds the intrinsic sub-pixel
accuracy. Further errors due to flow-induced distortion of the
particle image are estimated by comparing the imaging time $t_{\rm
im}$ with the time $t_{\rm f}$ required for the flow to displace
the particle over its own diameter. For a flow velocity
$\tilde{V}$ (in pixels per second), $t_{\rm
f}=2\tilde{a}/\tilde{V}$. We consider the particle significantly
distorted if $t_{\rm im}/t_{\rm f} \ge 0.1$. Hence, using
Eqs.~\ref{eq:ti2D},\ref{eq:ti3D}, the maximum velocities are:
\begin{equation}
\displaystyle{\tilde{V}^{max}_{2D} = 0.1 n f_{\rm scan},~~~~~~~~~\tilde{V}^{max}_{3D}=0.1 f \tilde{a}/\tilde{a}_{\rm z}}.
\label{eq:distortionlimit}
\end{equation}
Using the above parameters, a typical maximum velocity in 2D is
$V^{max} \simeq 500$~$\mu$ms$^{-1}$, while for 3D images with
$\tilde{a}/\tilde{a}_{\rm z}=1$, we obtain $V^{max} \sim
2~$~$\mu$ms$^{-1}$. In both cases, further improvement could be
achieved by removing the distortion prior to locating the
particles by correlating scanned images or lines at different
$z$~\citep{DerksJPCM04_coneplate}.

\subsection{Tracking algorithms}
\label{subsec:Tracking}

Merging particle coordinates from subsequent frames into
single-particle trajectories is the optimal route to analyse
colloidal dynamics. However, in some cases this is
difficult~\citep{BreedveldJCP2001_diffusionscales}, but
quantitative information may still be obtained via alternative
methods. For example, Breedveld {\it et
al.}~\citep{BreedveldJFlMech98,BreedveldJCP2001_diffusionscales,BreedveldJCP02_fulltensor}
measured the distribution of all possible displacement vectors in
images of tracer particles in dense non-Brownian suspensions. From
these distributions, the contribution due to cross correlations,
corresponding to vectors connecting the position of particle $i$
in one frame to that of particle $j$ in the next frame, could be
subtracted. From the resulting autocorrelation part of the
distribution, the full shear-induced self-diffusion tensor could
be obtained. While this method also has potential for analysis of
coordinate ensembles obtained from (confocal) microscopy during
flow, we will focus on a complete analysis of the particle
dynamics via tracking.

\subsubsection{Conventional Tracking }
\label{subsubsec:classicTracking}

The most widely used algorithm to track particles from ensembles
of coordinates in consecutive frames is that of Crocker and Grier
(CG,~\citep{CrockerGrierJColIntSc96_tracking}). It is based on the
dynamics of dilute non-interacting colloids. Given the position of
a particle in a frame and all possible new positions in the next
frame, within a `tracking range' $R_T$ of the old position, the
algorithm chooses the identification with the minimum mean squared
frame to frame displacement (MSFD). The algorithm has been used to
analyse particle dynamics in a wide variety of 2D and 3D images of
quiescent systems, see e.g.~\citep{weeksrev1,weeksrev2}. Its main
limitation is that, when the particle motion between frames is
excessive, misidentifications can occur. Such motion can be due to
diffusion, flow-advection or both.

In quiescent systems, the ability to track particles between
frames is limited by experimental constraints such as the
acquisition rate of the (confocal) microscope, image
dimensionality, particle size and concentration, and solvent
viscosity. After locating the particles, the relevant quantity
which relates to the tracking performance of the CG algorithm is
the root mean squared frame-to-frame displacement relative to the
mean interparticle distance~$\ell$. If particles move on average a
substantial fraction of $\ell$, then the algorithm starts to
misidentify them. This has recently been
quantified~\citep{isa_methods} by testing the CG algorithm on
computer-generated hard-sphere or hard-disk ensembles at different
concentrations. These tests showed that the algorithm can handle
larger mean squared displacements (MSDs) for more concentrated
systems, since the higher concentration prevents particles from
coming into close proximity of each other between frames.

The tests were also performed on computer-generated data in which
additional uniform or non-uniform motion was added, to study how
far the CG algorithm could be pushed beyond its original design
parameters. For uniform motion, CG tracking was as successful as
in the quiescent case for small drifts but failed for drifts of
the order of half the particle-particle separation. For
non-uniform (linear shear) flows with small strains between frames
the identification worked correctly, but large non-uniform
displacements caused major tracking errors.

\subsubsection{Correlated Image Tracking}
\label{subsubsec:correction}

Some important limitations to particle tracking in 2D or 3D images
with large drift or non-uniform motion (see also
\citep{XuReeveseRSciInstr04_trackingerrors}) can be overcome by
the method of Correlated Image Tracking (CIT), described in detail
in~\citep{isa_methods}. The main extension compared to
conventional (CG) tracking is that, prior to tracking, the time-
and position-dependent advective motion is obtained from a
PIV-type correlation analysis as in
Section~\ref{subsec:generalcorrel}. This advected motion is then
subtracted from the raw particle coordinates, shifting them to  a
`locally co-moving' (`CM') reference frame where the particles can
be tracked with the CG algorithm.

Since particle coordinates are distributed continuously, the
advection profile $\Delta {\bf R}(x_p,y_q,z_r,t_i)$ form the PIV
analysis is first interpolated to obtain a {\it continuous}
profile $\Delta {\bf R}(x,y,z,t_i)$. Using the latter, the
transformation of the position
$\mathbf{r}_k(t_i)=[x_k(t_i),y_k(t_i),z_k(t_i)]$ of particle $k$
in the laboratory frame to its position
${\mathbf{\bar{r}}}_k(t_i)$ in the CM reference frame is:
\begin{equation}
{\mathbf{\bar{r}}}_k(t_i) = {\mathbf{r}}_k(t_i) - \sum _{j=1}^{i}
{\mathbf{\Delta R( r_k(t_i)}},t_j) ,
\label{eq:general-advection-removal}
\end{equation}
with ${\mathbf{\Delta R}}( {\mathbf{r}}_k(t_i),t_j)$ the {\it
past} motion between frame $j$ and $j-1$, at the {\it current}
particle location ${\bf r}_k(t_i)$. In the CM frame, the average
particle motion is essentially zero over the entire image.
Therefore the classic CG algorithm tracks particles successfully
in this reference frame. The limitations on the CG tracking
performance are essentially the same as discussed in
Section~\ref{subsubsec:classicTracking} for quiescent systems.
Finally, the resulting trajectories can be restored in the
laboratory frame by inverting
Equation~\ref{eq:general-advection-removal}.

The main limitation to CIT originates from the failure of the
image correlation procedure when the {\it relative} particle
motion between frames is excessive, rather than from failure due
to large absolute shifts between images \citep{isa_methods}. This
apart, CIT has the same limitations of CG tracking with
significant failure for mean squared frame-to-frame displacements
in the CM reference of $\sim (0.3\ell)^2$ in 2D. Two direct
applications of CIT will be reviewed in
Section~\ref{subsec:disobserv}.

\section{Imaging of systems under deformation and flow}
\label{sec:observations}

Direct imaging remains the most detailed method for studying
structure and dynamics of particles in flow. This section reviews
the application of (confocal) imaging to study flowing
concentrated suspensions. We separately discuss disordered systems
and systems in which order is present before or as a result of
flow. In each case, the focus is on Brownian systems, but we
briefly review non-Brownian systems at the end of each subsection.

\subsection{Disordered systems}
\label{subsec:disobserv}

\subsubsection{Slow glassy flows}
\label{subsubsec:slowflow}

A significant motivation for recent developments in imaging
concentrated colloidal suspensions under flow is to investigate
the behaviour of samples that are so dense that structural
rearrangements are arrested in the quiescent (unsheared) state,
i.e. colloidal glasses \citep{sciortinoreview}. Elucidating the
mechanisms for the deformation and flow of colloidal glasses is
currently one of the `grand challenges' in soft matter science.
Much insight to this problem can be expected to come from the
detailed study of model systems. The simplest model systems are
concentrated hard sphere suspensions, which are glassy at volume
fractions $\phi \gtrsim 0.58$. (This claim has recently been
disputed \citep{luca}; but however this controversy eventually
resolves, it remains true that for almost all practical
rheological purposes, amorphous colloidal hard-sphere suspensions
behave as soft solids for $\phi \gtrsim 0.58$.)

The first time resolved studies of the flow of HS colloidal
glasses at single particle level have been described in
~\citep{BesselingPRL2007,BallestaPRL08,isa_methods}. Both the 3D
particle dynamics and global flow were investigated in steady
shear under various boundary conditions using both a planar shear
cell and `confocal rheoscope'
~\citep{BesselingPRL2007,BallestaPRL08} (geometries with uniform
stress in the flow gradient direction $z$). Standard fluorescent
PMMA-PHS particles ($a=850$~nm, $\phi=0.62$) in a charge-screened,
CHB-decalin mixture were used for observations at the particle
scale, measurements on a global scale employed fluorescent tracers
in a non-fluorescent host suspension.

We first focus on dynamics observed for rough boundary conditions.
Globally, these highly concentrated HS glasses shear
non-uniformly, showing rate-dependent shear localization
~\citep{BesselingPRL2007,BallestaPRL08,BesselingTBP}: at the
lowest applied rates $\dot\gamma_a$, the local shear rate
$\dot\gamma$ near (one of) the walls considerably exceeds
$\dot\gamma_a$ and vanishes smoothly into the bulk where the
system remains solid-like. The characteristic length, along $z$,
characterizing this shear rate variation starts from $\sim 25$
particle diameters and increases with the applied rate/stress,
i.e. the wall 'fluidization' propagates into the sample and
eventually leads to fully linear flow profile at the largest rate.

\begin{figure}
\includegraphics[width=0.3\textwidth,clip]{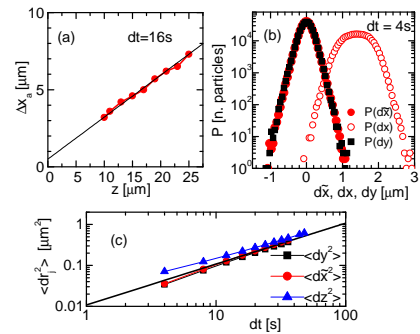}
\caption{Reprinted from~\citep{isa_methods}. 3D analysis of a
sheared glass at $\dot\gamma=0.019$~s$^{-1}$. (a)
(\Red{$\bullet$}) accumulated displacement $\Delta X_a(z,\Delta
t)$ (see Figure~\ref{fig:advectionsketch}) from image correlation
over $\Delta t=16$~s ($4$ frames). Line: linear fit giving an
accumulated strain $d\Delta X_a/dz=0.28$. (b) Distribution of
frame-to-frame displacements $P(\Delta x)$ and $P(\Delta y)$ from
Correlated Image Tracking with coordinates restored in the
laboratory frame. Also shown is $P(\Delta \tilde{x})$ of the
non-affine $x$-displacements, using Equation~\ref{eq:nonaffinedx}
and $\dot\gamma=0.019$~s$^{-1}$. (c) the (non-affine) MSD in the
three directions. Line: $\langle \Delta y^2(\Delta t) \rangle=2Dt$
with $D=5.4 \times 10^{-3}$~$\mu{\rm m}^2$s$^{-1}$.}
\label{fig:fastflow3D}
\end{figure}

The microscopic dynamics inside a shear band was then studied by
focussing on a $30\times30\times15$~$\mu$m$^3$ volume ($\sim 3000$
particles) the bottom layer of which was $\sim 10$ particles away
from the wall. A series of undistorted 3D snapshots of the entire
microstructure was acquired, from which 3D particle positions and
dynamics were obtained as described in Secs.~\ref{subsec:loc} and
\ref{subsec:Tracking}. Particle tracking using the new CIT method
allowed to study the flow for {\it local} shear rates up to
$\dot\gamma \simeq 0.05$~s$^{-1}$. The results are shown in
Figure~\ref{fig:fastflow3D}, where (a) presents the advected
motion $\Delta X(z,\Delta t)$ (from image correlation) for $\Delta
t=16$~s and a shear rate $0.02~$s$^{-1}$ \citep{isa_methods}. {\it
Local} velocity profiles such as these are linear on this scale
and extrapolate to zero within the resolution, showing the absence
of slip in this case. Figure~\ref{fig:fastflow3D}(b) shows the
distributions $P(\Delta x)$ and $P(\Delta y)$, which are
frame-to-frame particle displacements obtained from CIT. The
laboratory frame data for the displacements in the velocity
direction ($x$) show a large contribution from the advected
motion. To focus on the more interesting non-affine part of the
displacements, $\Delta \tilde{x}$, the $z$-dependent advected
motion was subtracted via:
\begin{equation}
\Delta \tilde{x}(t)=x(t)-x(0)-\dot \gamma \int_0^t z(t') dt', \label{eq:nonaffinedx}
\end{equation}
with $\dot\gamma$ measured form the data. The results for
$P(\Delta \tilde{x})$, Figure~\ref{fig:fastflow3D}(b), showed that
non-affine motion for $x$ and $y$ were very similar. Analysis of
the full 3D dynamics further revealed that it is indeed nearly
isotropic in all directions, as shown by the long time behaviour
of the mean squared displacements for $x$, $y$ and $z$ in
Figure~\ref{fig:fastflow3D}(c). The linearity of these curves at
long times also indicates that, after a sufficient number of shear
induced cage-breaking events, particle dynamics in the system
becomes diffusive, with long-time diffusion coefficients varying
by $<20\%$ in the three directions.

\begin{figure}
\includegraphics[width=0.3\textwidth,clip]{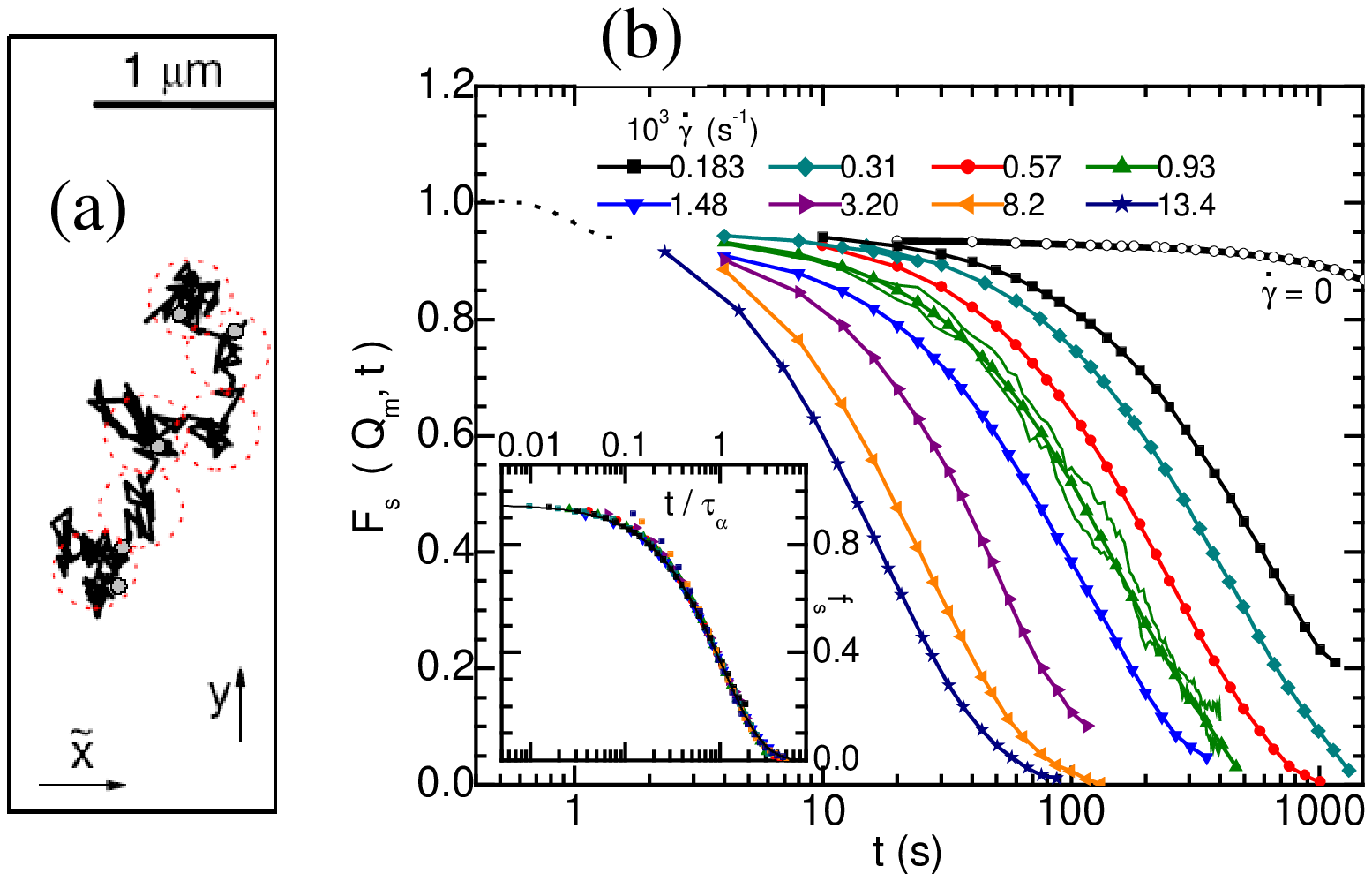}
\includegraphics[width=0.3\textwidth,clip]{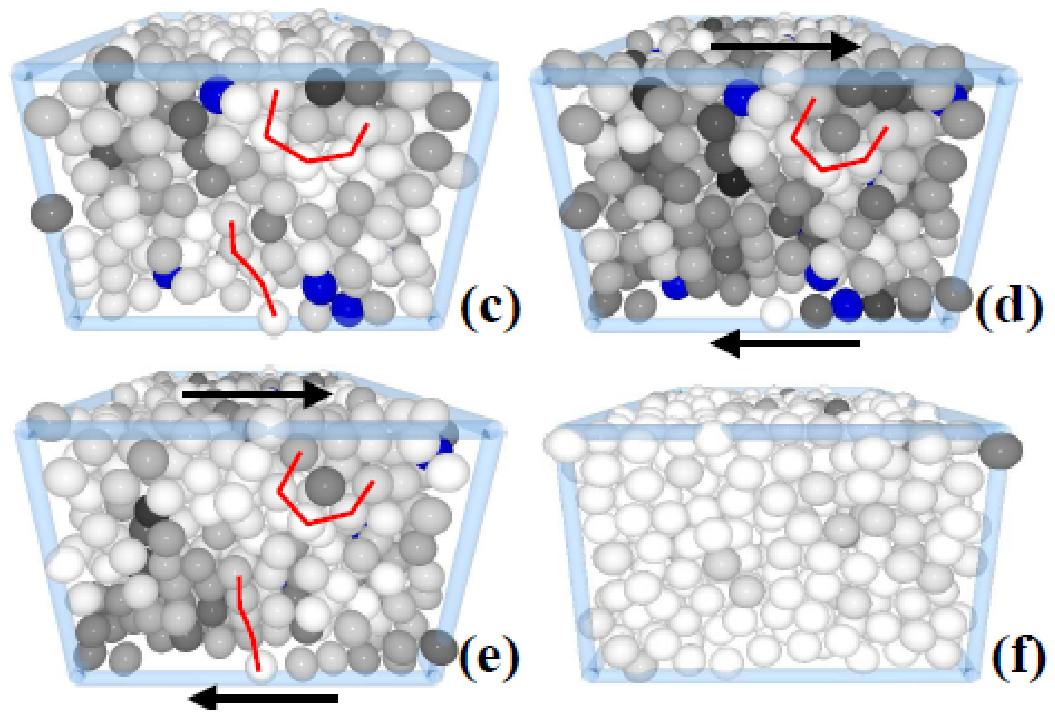}

\caption{Partly reprinted from~\citep{BesselingPRL2007}. (a)
Trajectory of a single particle over $800$~s in the
velocity-vorticity plane in the co-moving frame as defined in
Equation~\ref{eq:nonaffinedx} for a sheared glass at a local shear
rate $\dot\gamma=9.3\cdot 10^{-4}$~s$^{-1}$. (b) Selected
incoherent scattering functions $F_s(Q_m,t)$, with $\dot{\gamma}$
increasing from right to left. The dashed line schematises initial
relaxation. Inset: data collapse using $f_s(Q_m,t/\tau_{\alpha})$,
with the line showing that $f_s \propto \exp(-t/\tau_\alpha)$.
(c-f) Snapshots of the structure for
$\dot{\gamma}=0.0015$s$^{-1}$, (c) at $t_0=80$s after start of
imaging. The grey-scale measures the change in local environment
of each particle ($\propto C_6^i(t_0,dt)$) over the past $dt=40$s.
(d) at $t_0=120$s and $dt=80$s. (e) at $t_0=120$s and $dt=40$s.
Red lines show local deformations, arrows mark the shear
direction. (f) A quiescent glass, with $dt=200$s.}
\label{fig:track_correl}
\end{figure}

These experiments give information on individual plastic
cage-breaking events as well as rate-dependent structural
relaxation. The first phenomenon is illustrated by the non-affine
motion of a single particle in the $\tilde{x}-y$ plane,
Figure~\ref{fig:track_correl}(a). Intermittent jumps, reflecting
`plastic' breaking of the particle cages, are observed between
periods of `rattling' in which the cage is deformed. An average
accumulated strain of $\sim 10\%$ is found between these events,
in reasonable agreement with the yield strain obtained from bulk
rheology or light scattering
\citep{PetekidisJPCM04_creep+flow,PetekidisFarDisc03_DWSrheo}. The
changes in the particles' local environment during shear were also
studied via the change in the bond order parameters $Q_{6,m}^i$
for each particle $i$
\citep{WoldeFrenkelJChemPhys96_LJnucleationrate} over a time $dt$,
measured by the quantity $C_{6}^{i}(t,dt)$, where $C_6 =0$
reflects no change. Figure~\ref{fig:track_correl}(c)-(e) show
snapshots of the sheared microstructure at two times, where the
grey-scale is proportional to $C_{6}^{i}(t,dt)$. In
Figure~\ref{fig:track_correl}(c), the changes over $6\%$
accumulated strain are shown. Clusters of strong rearrangements,
termed `Shear Transformation Zones' (STZs) in earlier theoretical
studies \citep{FalkLangerPRE98_STZ}, can be observed, where the
local strain appears much higher than elsewhere. Over the
following $6\%$ accumulated strain,
Figure~\ref{fig:track_correl}(e), STZs appear in different
locations while the earlier ones remain essentially locked.

The average structural relaxation was examined via the incoherent
scattering function, $F_s(Q,t)=\langle
\cos(Q[y_i(t_0+t)-y_i(t_0)]) \rangle_{i,t_0}$ (using $Q =Q_m
\simeq 3.8a^{-1}$). As shown in Figure~\ref{fig:track_correl}(b),
the long-time dynamics is essentially frozen at rest
($\dot\gamma=0$). At short times and small shear rates, $F_s$
exhibits a plateau corresponding to caging at small accumulated
strain. At longer times, $F_s$ decays to zero due to repeated
cage-breaking events in line with the diffusive dynamics in
Figure~\ref{fig:track_correl}(c). This decay accelerates strongly
on increasing rate. When scaling time by the characteristic
($\alpha$) relaxation time $\tau_{\alpha}$, $F_s$ collapsed onto a
single exponential curve (see the inset), which directly confirmed
the theoretically predicted time-shear superposition principle.

The rate dependence of the inverse relaxation time and diffusivity
$D$ exhibited a non-trivial scaling: $D\sim 1/\tau_{\alpha} \sim
\dot\gamma ^{0.8}$. Such `power-law fluid' scaling contrasts the
yield-stress behaviour observed in the global rheology or
predictions from Mode Coupling Theory \citep{FuchsJRheol09}. (For
further discussion of these data in the context of Mode Coupling
Theory, see the article by Fuchs in this issue.) However, these
observations are consistent with a stochastic non-linear Langevin
equation treatment \citep{schweizer03a,schweizer03b}.  In this
approach to hard-sphere dynamics, particles become more deeply
trapped in effective free-energy (or, equivalently, entropic)
wells as the concentration increases beyond a certain critical
value; the effect of applied stress is to lower the entropic
barrier that particles have to surmount in order to escape
\citep{schweizer05}; the predicted shear-induced relaxation time
indeed shows power-law scaling with shear rates in an intermediate
regime with exponents close to $0.8$ \citep{schweizer08}.

\begin{figure}
\includegraphics[width=0.45\textwidth,clip]{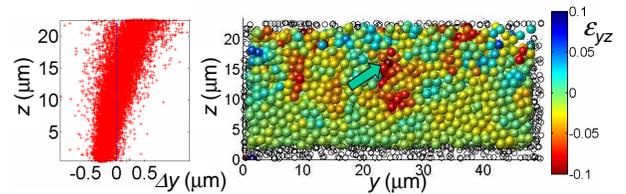}
\caption{Reprinted from ~\citep{SchallScience07_strainglass}, with
permission from AAAS. (A) Shear-induced displacements of particles
in a colloidal packing between $z = 0$ (the cover slide) and $z =
23~\mu$m after $50$ minutes of shear ($3\%$ average accumulated
strain). (B) Strain distribution and shear transformation zones in
a $7~\mu$m thick section in the shear-displacement gradient plane.
Particle color indicates the value of the local shear strain
$\epsilon_{yz}$ (see color scale), accumulated over $50$ minutes
of shear. The arrow indicates a shear transformation zone.
\label{fig:schallglass}}
\end{figure}

Subsequently, Schall {\it et al.} studied the behaviour of dense
colloidal packings under very small shear
deformations~\citep{SchallScience07_strainglass}. The packing was
formed by sedimentation of silica-spheres ($a=0.75~\mu$m) in a
lower density solvent, so that the volume fraction varied as
function of height. The shear geometry consisted of a movable,
coated bottom cover slide, while a metal grid positioned on top
confined the packing within a layer of $42~\mu$m ($\sim 30$
particle diameters). The  analysis focussed of the microscopic
strain variations resulting from either thermal fluctuations or
applied strain. From the thermal distribution of strain energies,
an estimate for the shear modulus was obtained, but volume
fraction gradients complicate the analysis. Large accumulated
local strains were observed at long times in the quiescent sample,
indicating aging. The behaviour under application of shear is
shown in Figure~\ref{fig:schallglass}. In this work, the total
accumulated strain was $\sim 3\%$ on average,
Figure~\ref{fig:schallglass}(a), well below the typical bulk yield
strain in colloidal hard sphere
glasses~\citep{PetekidisFarDisc03_DWSrheo,PetekidisJPCM04_creep+flow}.
Thus, it is probable that the data represent creep rather than
yielding and flow.

Figure~\ref{fig:schallglass}(b) shows a spatial map of the shear
strain $\epsilon_{yz}$ in a slice normal to the shear gradient.
The authors identified the localized regions with high strain with
STZs, the first published images of STZs in colloidal glasses.
Interestingly, the highlighted region revealed a four-fold
symmetric strain field just after its (earlier) formation,
consistent with theoretical predictions for a single  plastic
event. Schall {\it et al.} further identified the STZ-cores and,
using somewhat uncontrolled assumptions, calculated the STZ
formation energy, activation volume and the activation energy.
From this it was concluded that STZ formation was mainly thermally
activated, although some strain assistance was involved.

Images such as those shown in Figure~\ref{fig:schallglass}(b) and
\ref{fig:track_correl}(c)-(e) show that direct imaging is a very
powerful method for studying inhomogeneous local responses to
applied stress. In particular, the discovery of STZs in model
colloids means that the physics of these systems may have
relevance beyond soft matter physics, e.g. in the study of
large-scale deformation of metallic glasses \citep{LangerReview}.

\begin{figure}
\includegraphics[width=0.45\textwidth,clip]{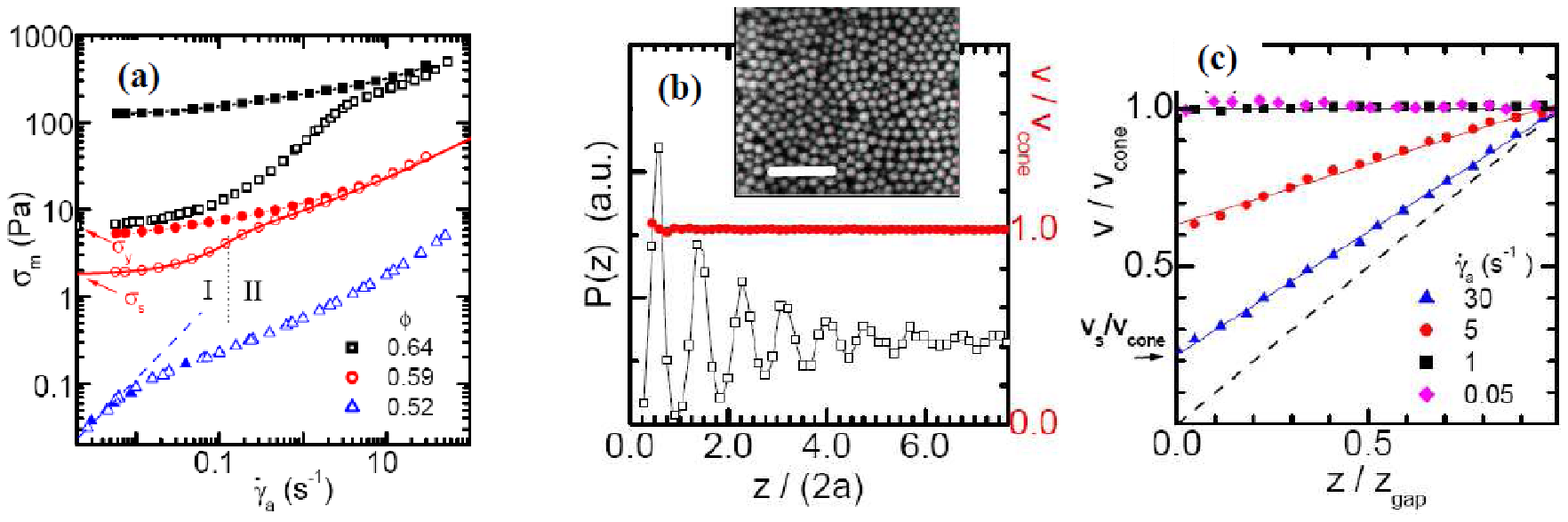}
\caption{Reprinted from~\citep{BallestaPRL08}. (a) Measured stress
$\sigma_m$ versus applied rate $\dot\gamma_a$ for particles with
$a=138$~nm, at different volume fractions for coated
($\blacksquare$,\Red{$\bullet$},\Blue{$\blacktriangle$}) and
un-coated geometries ($\square$,\Red{$\circ$},\Blue{$\triangle$}).
Dashed line: linearly viscous behaviour. The full line is a flow
curve predicted using the model in~\citep{BallestaPRL08}. Dotted
line: applied rate at which yielding starts at the cone-edge. (b)
Density profile $P(z)$ ($\square$), from 3D-imaging of the
$a=652$~nm system at $\phi=0.63$, during slip at $r=2.5$~mm and
$\dot\gamma_a=0.01$~s$^{-1}$. (\Red{$\bullet$}) Corresponding
velocity profile from particle tracking, showing full plug flow.
Inset: Slice of one of the 3D-images, showing the first colloid
layer. Scale bar: $10$~$\mu$m. (c) Velocity profiles for the
$\phi=0.59$ data in (a), in units of the cone velocity $v_{\rm
cone}=\dot\gamma_a \theta r$, as function of the normalized hight
$z/z_{\rm gap}=z/\theta r$, for various applied rates
$\dot\gamma_a$ at $r=2.5$~mm. The arrow marks the slip velocity
for $\dot\gamma_a=30$~s$^{-1}$; the dashed line is the behaviour
without slip $v=\dot\gamma_a z$ as observed for $\phi < \phi_{\rm
G}$.} \label{fig:slipplots}
\end{figure}

Particulate suspensions and other concentrated soft materials
(e.g. emulsions) often exhibit wall slip during flow along smooth
boundaries. This is important in practical applications, and
strongly relevant when interpreting bulk flow measurements. Slip
in particulate suspensions has been seen mostly for high
P\'{e}clet numbers (non-Brownian suspensions), both in
solid-~\citep{Yilmazer,KalyonJRheo05_slip} and
liquid-like~\citep{JanaJRheo95} systems. In a recent study,
Ballesta {\em et al.} addressed the effect of Brownian motion and
the glass transition on slip in hard-sphere colloids using
confocal imaging and simultaneous rheology~\citep{BallestaPRL08}
(Section~\ref{subsubsec:rotational}). The main results are shown
in Figure~\ref{fig:slipplots}. The rheology for rough walls in
Figure~\ref{fig:slipplots}(a) shows the traditional change from a
shear thinning fluid ($\phi=52\%$) to a yielding solid
($\phi=59\%$) associated with a glass transition at $\phi \sim
58\%$~\citep{fuchs1,PetekidisJPCM04_creep+flow}. For a smooth wall
and index-matching solvent (which prevents van der Waals
attraction and sticking of particles to the glass) the results for
a dense liquid remained unchanged, indicating no slip. This is
confirmed by imaging (Figure~\ref{fig:slipplots}(c), dashed line)
and also agrees with earlier velocity profile measurements for
dense colloidal liquids~\citep{DerksJPCM04_coneplate}. However,
for the colloidal glass, a slip branch developed in the flow curve
at small rates. This branch was described by a Bingham-form,
associated with solid-body motion of the suspension along the
glass plate. This `solid' structure during slip was shown to
extend down to the first particle layer,
Figure~\ref{fig:slipplots}(b) and the inset, giving microscopic
insight in the 'Bingham' slip-parameters. The transition from pure
slip at low rates to yielding at high rates was also probed, see
Figure~\ref{fig:slipplots}(c). This transition depended on the
local gap, i.e. the radial position $r$, illustrating the
possibility of non-uniform ($r$-dependent) stress in a cone-plate.

Interestingly, this slip behaviour of hard-sphere glasses is
different in nature from that found earlier by Meeker {\it et al.}
in jammed systems of emulsion
droplets~\citep{MeekerPRL04_rheoslip}. There, a non-linear
elasto-hydrodynamic lubrication model, appropriate for deformable
particles, could quantitatively account for their observations. It
therefore appears that while slip is ubiquitous, the mechanism for
its occurrence can be highly system dependent.

\subsubsection{Fast channel flow of fluids and pastes}

A different regime of flow in hard-sphere suspensions, either in
dense fluids or glasses, occurs when the shear rate becomes
comparable to the rate for thermal relaxation of the particles
within their cages, which is of the order of the inverse Brownian
time. In these situations jamming or shear thickening may start to
occur, while in geometries with non-uniform stress, pronounced
shear-induced migration may take place. Here we describe recent
studies of imaging in this regime, focussing on the specific
geometry of capillaries and micro-channels.

The effects of shear-induced migration in intermediate volume
fraction hard-sphere suspensions ($\phi < 0.35$) was studied by
Frank \textit{et al.}~\citep{frank} and Semwogerere \textit{et
al}.~\citep{semwogerere1}. In these experiments, confocal imaging
was used to measure both velocity and concentration profiles,
$\phi(y)$, of colloids across $50 \mu$m$\times 500 \mu$m glass
channels. Flow velocities up to $8~$mms$^{-1}$ were probed,
corresponding to P\'{e}clet numbers up to $\sim 5000$. The
gradient in shear rate $\dot\gamma(y)$ causes particles to migrate
from the boundaries (with large $\dot\gamma$) to the center of the
capillary (small $\dot\gamma$), resulting in development of a
steady state concentration profile beyond a certain 'entrance
length'. Frank {\it et al.} found profiles (transverse to the
flow) where the concentration in the center increased considerably
with the average volume fraction or on increasing flow rate
(P\'{e}clet number, Pe). The data were interpreted via a model
which included a shear rate and local volume fraction dependence
of normal stresses in the sample. The rapid rise of these stresses
with $\phi$ and Pe was responsible for the observed behaviour.
Semwogerere \textit{et al}.~\citep{semwogerere1} studied the
entrance length over which the flow became fully developed and
found that it increases strongly with Pe, in marked contrast to
non-Brownian flows for which it is flow-rate independent
\citep{NottBradyJFM94}. The entrance length also decreased with
increasing volume fraction and the data could be successfully
described within their model.

Few imaging studies have been carried out in higher concentration
suspensions. Haw~\citep{HawPRL04_jamming} observed intriguing
behaviour in traditional microscopy studies of PMMA suspensions
($\phi \gtrsim 55\%$) sucked into mm-sized capillaries with a
syringe. Strikingly, the volume fraction of the suspension in the
capillary was as much as $5\%$ lower than the initial
concentration, depending on the particle size. Such unexpected
``self-filtration'' was attributed to jamming of the particles at
the capillary entrance, leading to higher flow rate of the pure
solvent under the applied pressure difference.

\begin{figure}
(a)\includegraphics[width=0.3\textwidth,clip]{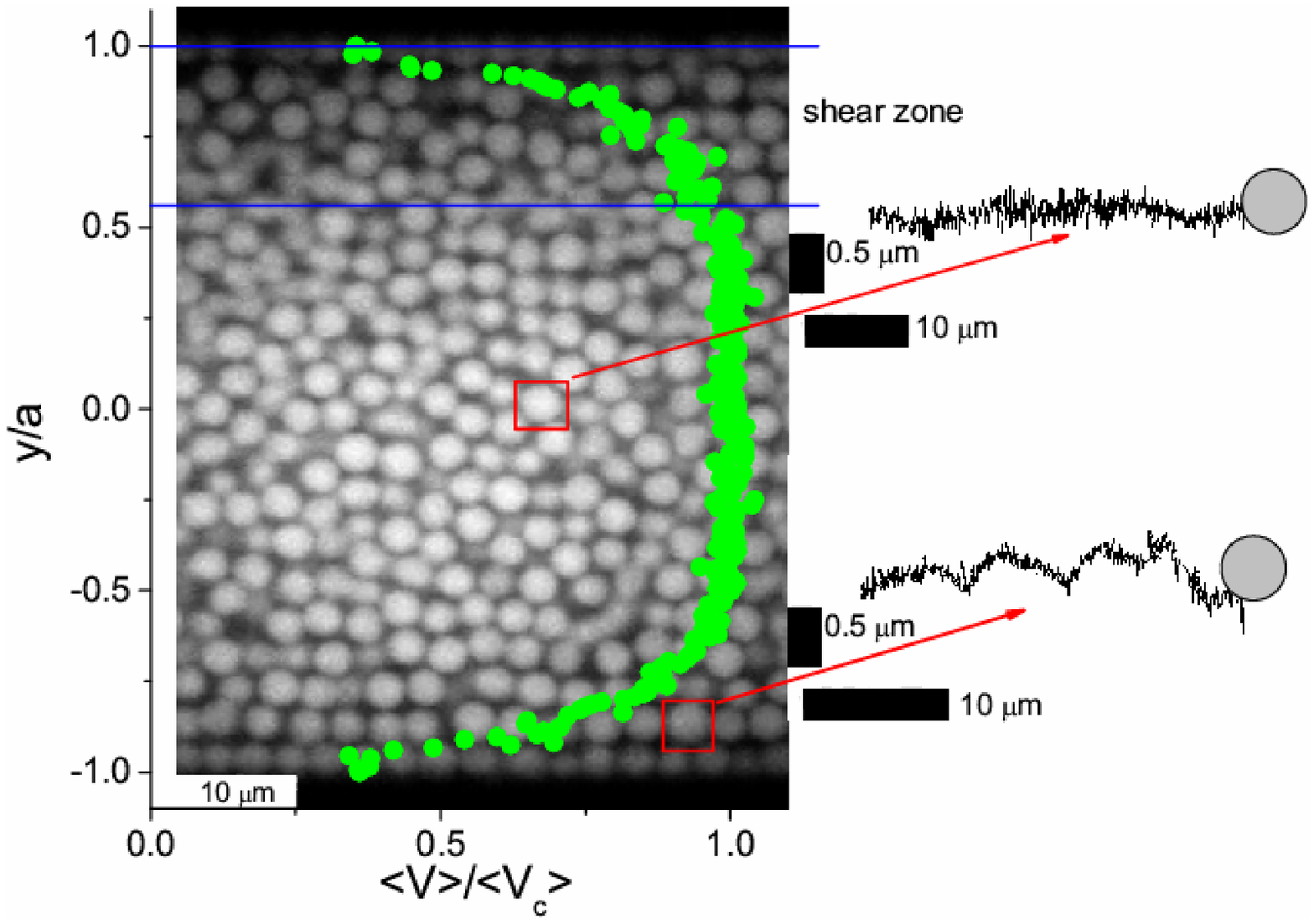}(b)
\includegraphics[width=0.3\textwidth,clip]{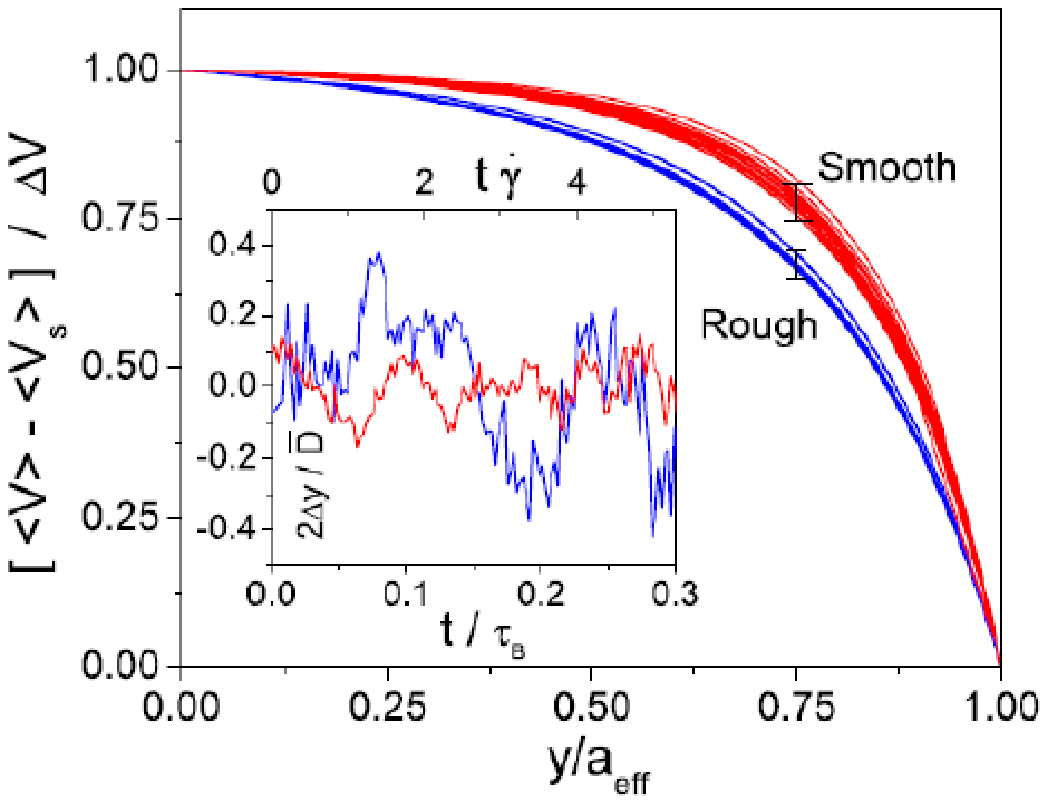}
\caption{(a) Velocity profile of a 63.5\% volume fraction
suspension of PMMA hard-spheres (radius $a = 1.3 \mu$m) in a
square, glass micro-channel (side $2 b=50 \mu$m). The velocity
$\langle V \rangle$ is normalized by the centerline velocity
$\langle V_{\rm c} \rangle$. The profile is overlaid onto a
confocal image of the suspension in the channel. The velocity
profile consists of a central, unsheared plug and of lateral shear
zones, whose width is  highlighted by the horizontal blue lines.
The right side of the image shows two examples of particles
trajectories in the plug and in the shear zone respectively. The
$x$ and $y$ scales of the  trajectories are different and the
colloids are not drawn in scale. (b) Reprinted from~\citep{isa2}.
Normalised velocity profiles at $z = 17$~$\mu$m versus $y/b_{\rm
eff}$ ($b_{\rm eff}= b-a$) for a wide range of flow rate. For
clarity, fits to the data rather than the raw data are shown; the
error bars show the spread in the measurements. Inset: normalized
transverse fluctuations of individual particles in the shear zone
for the two boundary conditions, versus normalized time $t/\tau_B$
or {\it local} accumulated strain $t \cdot
\partial \langle V(y) \rangle/\partial y=t \cdot \dot{\gamma}(y)$, top axis.}
  \label{fig:luciochannel}
\end{figure}

Motivated by these experiments, Isa {\em et al.} conducted further
studies of flows in similar geometries at the single particle
level using confocal microscopy ~\citep{isa2}. The system
consisted of a hard-sphere suspension (PMMA spheres, radius $1.3
\pm 0.1$~$\mu$m) at nearly random close packing, `a paste', in a
20-particle-wide square capillary. The motion of individual
colloids was tracked via Correlated Image Tracking and velocity
profiles were measured in channels with both smooth and rough
walls. Despite the colloidal nature of the suspension, significant
similarities with granular flow ~\citep{GranularReview,MiDi} were
found.

The bulk flow curve of the system studied by Isa {\em et al.}
fitted a Herschel-Bulkley (HB) form for a yield-stress
fluid~\citep{PetekidisJPCM04_creep+flow} at small to moderate flow
rates. The velocity profile predicted from a HB constitutive
relation consists of a central, unsheared ``plug'' and shear zones
adjacent to the channel walls, which fits qualitatively with what
was observed, Figure~\ref{fig:luciochannel}(a). At the same time,
however, the size of the central plug is expected to decrease with
increasing flow rate, vanishing beyond some critical velocity
above which the yield-stress fluid is shear-melted everywhere (see
e.g. ~\citep{You05,Steffe}). But Isa {\em et al.} found that width
of the central plug was independent of applied flow rate,
Figure~\ref{fig:luciochannel}(b). This behaviour is analogous to
that observed in the pipe flow of dense dry
grains~\citep{MiDi,Pouliquen1}. The data for the colloidal flow
could be captured by a theory of stress fluctuations originally
developed for the chute flow of dry granular
media~\citep{ShearZoneOrigins,Pouliquen1}. Presumably, this model
can be successfully applied to nearly-close-packed colloidal
systems because at high flow rates, inter-particle contacts rather
than Brownian motion dominate. Direct imaging showed indeed that
the trajectory of a particle in the shear zone was determined by
'collisions' with neighbouring particles,
Figure~\ref{fig:luciochannel}. Moreover, the presence of rough
boundaries enhanced such fluctuations, causing wider shear zones.

In more recent work on the same system ~\citep{isa_oscill}, using
smooth walls, it was shown that confinement can induce
instabilities in the flow. Upon decreasing the width of the
channels from $\sim 40$ to $\sim 20$ particle diameters, the flow
developed oscillations above a threshold applied flow rate. Such
oscillations consisted of cyclical jamming and un-jamming of the
suspended particles and led to filtration effects similar to those
reported by Haw~\citep{HawPRL04_jamming}. Single-particle imaging
was used to demonstrate the presence of a concentration profile
across the channel, which was well correlated with the local
velocity profile.

\subsubsection{Non-Brownian suspensions}

Other recent imaging experiments address the flow of non-Brownian
suspensions (i.e. $\mbox{Pe} \rightarrow \infty)$ at the single
particle level. In a series of
papers~\citep{BreedveldJFlMech98,BreedveldPRE2001,BreedveldJCP2001_diffusionscales,BreedveldJCP02_fulltensor},
Breedveld {\it et al.} investigated the steady shear-induced self
diffusion of non-Brownian spheres. They used PMMA particles
(radius $a \sim 45~\mu$m) in a RI and density-matching mixture
(water, zinc chloride, and Triton X-100, viscosity $\sim
3.4~$Pa~s) and studied suspensions with volume fractions ranging
from $20\%$ to $50\%$. By studying the correlated motion of a
small fraction of the particles which had been dyed
(Section~\ref{subsec:Tracking}), the self and `off-diagonal' mean
squared displacements (MSDs) could be extracted without the use of
explicit particle tracking.

For these non-Brownian suspensions, the long-time shear-induced
MSD depend on accumulated strain $\dot\gamma \Delta t$ only. The
`long-time' diffusion for $\dot\gamma \Delta t
> 1$ is due to the chaotic nature of multiple particle
collisions. The associated self-diffusion constants $D_{xx}$,
$D_{yy}$, $D_{zz}$ (in the velocity, vorticity and gradient
direction respectively) were found to be anisotropic with
$D_{zz}/D_{yy} < 2$, while $D_{xx}$ was almost an order of
magnitude larger. The data were in qualitative agreement with
hydrodynamic theories and simulations, although quantitative
discrepancies remained. In contrast, for the {\it slowly} sheared
glasses in
Section~\ref{subsubsec:slowflow}~\citep{BesselingPRL2007}, which
exhibited a non-trivial rate dependence, $D_{yy}/(a^2 \dot\gamma)$
scales as $\dot\gamma^{-0.2}$, ranging from $0.7$ at low rates to
$\sim 0.3$ at the largest rates, exceeding the result for
non-Brownian suspension. A last notable observation in the
non-Brownian system in~\citep{BreedveldJCP2001_diffusionscales}
was a regime of reduced diffusion for small accumulated strains
$\dot\gamma \Delta t \simeq \Delta \gamma_{cage} \ll 1$. While its
nature remains unclear, it was shown that the typical strain
$\Delta \gamma_{cage}$ roughly corresponded to the affine
deformation $\delta/a$ at which particles come in direct contact
(on average), which varies with volume fraction as
$(\phi_{rcp}/\phi)^{1/3}-1$, and vanishes as the average
surface-surface separation between particles on approaching close
packing.

In a very recent study by Wang {\it et al.} \cite{WangMaksePRE08},
the three-dimensional motion and diffusion of heavy tracer
particles in an immersed, density matched granular packing ($\phi
\sim 0.6$) in a cylindrical Couette-cell was studied. Among many
other remarkable results, the data in \cite{WangMaksePRE08} showed
direct evidence that $D \propto \dot\gamma$ in non-Brownian
systems. Interestingly, the value for shear diffusion in the
vorticity direction, $D/(\dot\gamma a^2) \simeq 0.1$ was close to
the values found by Breedveld for slightly lower density.

Another study on non-Brownian
suspensions~\citep{PineBradyNat05_irreversibility} focussed on a
transition from reversible to irreversible particle dynamics in
oscillatory shear. While the creeping (Stokes) flow equations for
simple shear flow are in principle reversible, particle roughness,
three particle collisions or repulsive forces (ignoring Brownian
motion) may render the dynamics irreversible. Pine {\it et al.}
showed that there exists a volume-fraction-dependent critical
strain amplitude beyond which the dynamics become diffusive. In a
follow-up study~\citep{CortePineNatPhys08_randomorganization},
they demonstrated that the state below the critical strain is in
fact an `absorbing state', where the particles self-organize into
a structure where collisions are avoided. They also demonstrated
an increase in the dynamic viscosity of the suspension on crossing
the critical strain amplitude, as was also observed previously by
Breedveld {\em et al.}~\citep{BreedveldJCP2001_diffusionscales}.
An interesting question is whether any relation exists between the
absorbing states measured in these non-Brownian systems and the
behaviour below the yield strain in colloidal glasses.

To date, flow studies of Brownian and non-Brownian systems have
largely been conducted independently. It is clear that much can be
gained by detailed comparison of such studies~\citep{coussotPRE}
and that single-particle imaging can provide crucial insights in
this exercise~\citep{isa2,isa_vprofilesbig}.

\subsection{Ordered systems and ordering under flow}
\label{subsec:order}

Hard spheres colloids undergo an entropy driven fluid-crystal
phase transition when the suspension's volume fraction reaches
$\phi_f = 0.494$. This behaviour was first confirmed
experimentally using quasi-monodisperse hard-sphere-like PMMA
colloids \citep{pusey1}. Static light scattering
\citep{puseyrandomstack} shows (and later, conventional microscopy
\citep{Bristol97} confirms) that the resulting colloidal crystals
consist of hexagonally-packed layers more or less randomly stacked
on top of each other; i.e., these crystals have a mixture of
face-centered cubic (fcc) and hexagonal close packed (hcp)
structures. Starting with a paper in 1998 \citep{henderson98}, van
Megen and co-workers have published a series of studies in the
kinetics of crystallization in PMMA suspensions using
time-resolved dynamic and static light scattering, investigating
in particular the effect of polydispersity \citep{bryant06} in
some detail.

Gasser and co-workers~\citep{gasser} pioneered the use of confocal
microscopy to study the kinetics of colloidal crystallization.
Using buoyancy-matched PMMA particles, they observed nucleation
and growth of colloidal crystals with single-particle resolution.
These authors identified the size of critical nuclei, 60-100
particles, in rough agreement with computer
simulations~\citep{auer}, and to measure nucleation rates and the
average surface tension of the crystal-liquid interface. For
completeness we mention that Palberg and coworkers have pioneered
the use of non-confocal microscopy methods to image ordering in
highly-deionized suspensions \citep{BiehlPalbergEPL04,palberg},
which crystallise at very low volume fractions ($10^{-3}$ or
lower).

The relative mechanical weakness of colloidal crystal compared to
their atomic and molecular counterparts means that modest shear
will have large, non-linear effects on crystallisation as well as
crystal structure. In the case of hard-sphere-like PMMA colloids,
these effects have been studied using light scattering
\citep{ackerson} since almost immediately after the discovery of
crystallisation in this system \citep{pusey1}. Much more recently,
confocal microscopy has been used to image shear effect in PMMA
colloids. Below, we focus on reviewing such studies, but we note
that imaging studies of crystallisation in other colloids have
also been performed (see, e.g., \citep{BiehlPalbergEPL04}). To
date, there have been few imaging studies of shear effects on
ordering and ordered states in non-Brownian suspensions. We review
briefly one series of such work at the end of this section.

\subsubsection{Hard-sphere colloids}
\label{subsubsec:Brownianordering}

It is known from static light scattering~\citep{ackerson},
diffusing wave echo spectroscopy \citep{haw3} and conventional
optical microscopy ~\citep{haw2} that the application of
oscillatory shear to dense colloidal fluids or glasses can drive
crystallisation. Rheological studies~\citep{Koumakis} show that
shear-crystallised samples have lower elastic and viscous moduli
than their glassy counterparts at the same volume fraction, in
agreement with the enhanced effective 'cage' volume in the ordered
case, leading to smaller entropic stiffness.

\begin{figure}
\includegraphics[width=0.45\textwidth,clip]{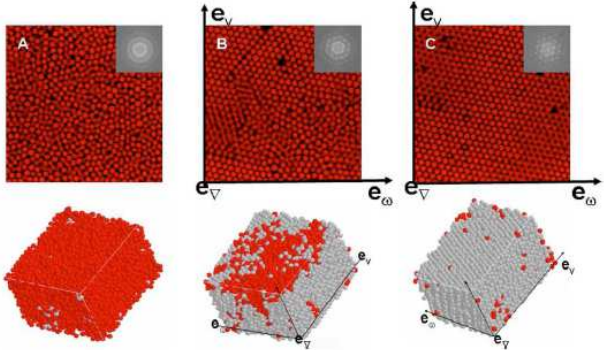}
\caption{Reprinted with permission from
\citep{SolomonJCP06_shearcrystalCF}. Copyright [2006], American
Institute of Physics. Confocal images (top row, with insets
showing the Fourier transform) and reconstructed 3D stacks (bottom
row) showing the time dependent effect of oscillatory shear
(frequency = 3~Hz, strain amplitude = 1) on a 48\% volume fraction
PMMA suspension.  Time progresses from left to right: quiescent
(24 hours after preparation, A), after 2 minutes (B), and after 5
minutes (C) of shear. In (B) and (C), unit vectors (e) in the the
velocity (v), vorticity ($\omega$) and shear gradient ($\nabla$)
directions are shown. The light grey particles in the 3D
renderings (bottom row) represent those with a high degree of
local crystalline ordering.
  \label{fig:solomon}}
\end{figure}

A detailed confocal imaging study of shear-induced crystallisation
in hard-sphere colloids was performed recently by Solomon and
Solomon~\citep{SolomonJCP06_shearcrystalCF}. They imaged crystals
formed under an oscillatory shear field at a particle volume
fraction of 52\% in slightly-charged PMMA particles (diameter
1.15~$\mu$m, polydispersity 4\%) in a planar shear cell
(Section~\ref{subsubsec:parallelplates}). Particles were
identified using the conventional Crocker and Grier algorithm but,
due to the low acquisition rate (0.6 frames/second), imaging was
only possible after the cessation of shear. Consistent with
previous work \citep{ackerson,haw2}, they observed ordering of the
initially amorphous suspension into close-packed planes parallel
to the shearing surface, Figure~\ref{fig:solomon}. Upon increasing
the amplitude of the applied oscillatory strain, $\gamma$, the
close-packed direction of these planes was observed to shift from
an orientation parallel to the vorticity direction to one parallel
to the flow direction, and the quality of the layer ordering
decreased. In addition, they studied shear-induced stacking faults
and reported their three dimensional structure. For large STRAIN
amplitudes ($\gamma = 3$), ordering in the flow-vorticity plane
only persisted for 5 to 10 layers, while in the gradient direction
the crystal consists of alternating sequences of fcc and hcp
layers.

\begin{figure}
\includegraphics[width=0.45\textwidth,clip]{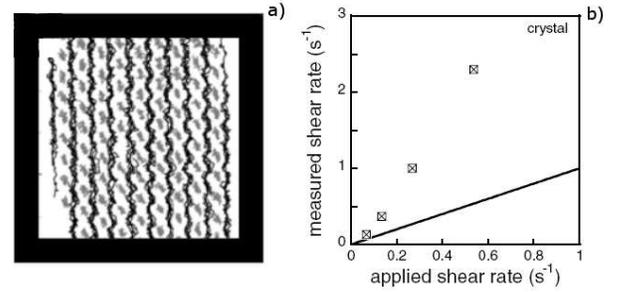}
\caption{Reprinted with permission from
~\citep{DerksJPCM04_coneplate}, copyright [2004], Institute of
Physics Publishing (a) Superposition of particle tracks in the
zero-velocity plane plus one adjacent plane in a sheared colloidal
crystal, showing zig-zag motion. The image size is 18.75 $\mu$m
$\times$ 18.75 $\mu$m (256 $\times$ 256 pixels). The colloids have
a diameter of 1.50 $\mu$m. (b) Measured shear rate versus applied
shear rate for the colloidal crystal (points), the continuous line
shows equality of these two rates. \label{fig:derks_track}}
\end{figure}

Once formed, whether it be under quiescent or sheared conditions,
colloidal crystals show a complex response to applied stress. In
particular, they exhibit banding under simple shear. We have
already discussed flow localization in colloidal glasses in
Section~\ref{subsubsec:slowflow}. The first evidence for the
occurrence of this phenomenon in ordered structures was reported
in a study of hard-sphere colloidal crystals under steady shear by
Derks \textit{et al.}~\citep{DerksJPCM04_coneplate}. They studied
the velocity profiles and single-particle dynamics via correlation
techniques and particle tracking in a cone-plate system
(Section~\ref{subsubsec:rotational}) around the zero-velocity
plane. Since particles were locked in their lattice positions and
their motion was limited, they could be tracked using the
conventional Crocker-Grier algorithm. The close-packed direction
in hexagonally-ordered planes was observed to be parallel to the
velocity, in agreement with high strain amplitude oscillatory
shear results~\citep{haw2,SolomonJCP06_shearcrystalCF}. The study
also showed direct evidence that colloids in adjacent layers move
in a zig-zag fashion, Figure~\ref{fig:derks_track}(a), as
previously inferred from light scattering~\citep{ackerson2} and
conventional imaging~\citep{haw2}. The velocity profile measured
over the accessible $z$ range were linear, but the measured shear
rates were much larger than the applied rates,
Figure~\ref{fig:derks_track}(b), indicating considerable shear
banding in the sample. In a more recent study
\citep{DerksBlaaderenSoftMatter09_shearxtal}, Derks {\it et al.}
analysed in more detail the particle dynamics in the
flow-vorticity plane of a sheared crystal. They evidenced
increasing fluctuations in the crystal layers on increasing rate,
eventually leading to shear melting when the short-time mean
square displacements due to shear attained a `Lindemann' value of
$\sim 12\%$ of the particle separation.

More direct evidence of shear banding in colloidal crystals under
{\it oscillatory} shear was found by Cohen and co-workers
\citep{Cohen_oscill_shear_2006}. In this confocal microscopy
study, the oscillatory motion across the gap was measured at
various applied amplitudes and frequencies. For low applied
deformation, $\gamma_{\rm app}$, the authors claim that the
crystal is linearly strained at all frequencies, $f$. At higher
$\gamma_{\rm app}$, the crystal yields, and two different regimes
are observed. At low frequencies, a high-shear band appears close
to the upper (static) plate. As the applied strain and frequency
increase, the band becomes larger until the whole gap is sheared
at $f >  3$~Hz. Moreover, large slip is observed in all cases,
with the measured strain always below the applied one. Cohen {\em
et al.} proposed a model based on the presence of two coexisting,
linearly-responding phases to explain their observations. This
model is attractive for its simplicity, but questions remain, in
particular concerning the lack of bulk strain for small-amplitude,
low-frequency data and some significant deviations between the
modelled and measured profiles.

Simple shear is, of course, a highly idealised deformation
geometry. Motivated partly by the possibility of shedding light on
the deformation, fatigue and fracture of metals in more
`real-life' situations, Schall and co-workers carried out a
confocal imaging study of a more complex form of deformation:
indentation in defect-free colloidal crystals grown by slowly
sedimenting silica particles on templated surfaces~\citep{
Schall_crystal_indentation_2006}. Indentation (or nano-indentation
in the case of metallic crystals) consists in pressing a sharp tip
(diamond for metals, and a needle for colloidal crystals) against
a surface to probe its mechanical properties by relating the
applied pressure and the measured penetration. It is also a method
for the controlled introduction of defects into a crystal.

Schall \textit{et al.} indented colloidal crystals using a needle
with an almost hemispherical tip of diameter 40$\mu$m, inducing a
strain field in which the maximum shear strain lies well below the
contact surface. The tip diameter, particle radius and crystal
thickness in their experiments were chosen to be similar to
parameters in typical metallic nano-indentation experiments. The
authors discussed their observations using a model that addresses
the role played by thermal fluctuations in the nucleation and
growth of dislocations.

\begin{figure}
\includegraphics[width=0.45\textwidth,clip]{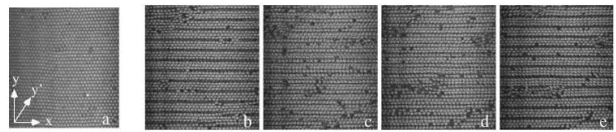}
\caption{Reprinted with permission
from~\citep{Cohen_confined_crystal_2004}, copyright 2004 by the
American Physical Society. Structure of a sheared suspension
(diameter 1.42~$\mu$m) with strain amplitude $\gamma_0=0.38$,
frequency $f=30$~Hz and $\phi=0.61$. The plate moves in the $x$
direction. (a) Confocal micrograph of a sheared suspension forming
hcp layers when the gap between the plates $D=80~\mu$m. (b)--(e)
images of the suspension in the buckled state. The gap is set
slightly below the height commensurate with confinement of four
flat hcp layers. (b) an $xy$ image slice of the suspension near
the upper plate. (c)-–(e) show slices that are, respectively,
$1.3$, $2.6$, and $3.9~\mu$m below the slice in (b).}
\label{fig:cohenconfined}
\end{figure}

We have already discussed confinement effects in the channel flow
of colloidal glasses. Such effects are also seen in hard-sphere
colloidal crystals sheared between parallel plates. Cohen {\em et
al.} ~\citep{Cohen_confined_crystal_2004} found that when the
plate separation was smaller than 11 particle diameters,
commensurability effects became dominant, with the emergence of
new crystalline orderings. In particular, the colloids organise
into ``$z$-buckled'' layers which show up in $xy$ slices as one,
two or three particle strips separated by fluid bands,
Figure~\ref{fig:cohenconfined}. By comparing osmotic pressure and
viscous stresses in the different particle configurations, the
cross-over from buckled to non-buckled states could be accurately
predicted.

\subsubsection{Non-Brownian suspensions}
\label{subsubsec:nonBrownianordering}

\begin{figure}
\includegraphics[width=0.45\textwidth,clip]{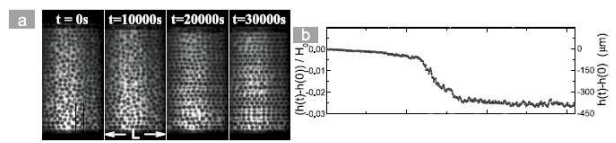}
\caption{Reprinted with permission from~\citep{TsaiPRL03},
copyright 2003 by the American Physical Society. (a) Time
evolution of the structure of an index matched suspension of glass
beads in ~\citep{TsaiPRL03}, imaged in the flow-gradient plane
after the start of shearing at $t=0$. The observed layering is
parallel to the flow direction. (b) Fractional volume compaction
vs time, based on the change of height of the upper driving
boundary. The actual change in microns is indicated at the right.}
\label{fig:tsai}
\end{figure}

Shear induced structuring has been predicted in the flow of
non-Brownian suspension (e.g.~\citep{sierou}), but to date there
have been few imaging studies. In a series of experiments by Tsai
and
co-workers~\citep{TsaiPRL03,TsaiPRE04_imaging+torque,TsaiPRE05_imaging},
laser sheet imaging was employed to study the dynamics of confined
glass sphere packings (diameter $600$~$\mu$m) in an index-matching
(non-density matching) solvent in an annular shear cell with load
applied from the top shearing surface. Using simultaneous
measurements of the volume (and torque in
\citep{TsaiPRE04_imaging+torque}), they discovered a transition
from a disordered packing to an ordered packing under steady
shear, Figure~\ref{fig:tsai}(a), accompanied by compaction steps,
Figure~\ref{fig:tsai}(b), and reduction of the shear force. They
observed strongly non-linear shear profiles with a rapid reduction
of the rate away from the top driving plate, varying significantly
with the order of the packing. Due to these non-linear profiles,
the delay observed before the ordering transition was strongly
dependent on the layer thickness for fixed boundary speed, since
global ordering requires an accumulated strain $\Delta \gamma \sim
1$ through the entire sample.

\section{Conclusion and outlook}

Recent advances in imaging and data analysis techniques have
enabled time resolved imaging and tracking of individual particles
in colloidal systems under flow. We have reviewed these
techniques, as well as examples of their application, primarily on
concentrated systems of hard-sphere colloids. Despite what has
been achieved, we believe that this area is only in its infancy.
Results already obtained raise many more questions than they
answer. Many phenomena and systems remain little explored, and
existing results invite further investigation using imaging
techniques. For example, computer simulations of crystal
nucleation \citep{frenkelshearednucleation} as well as proposed
explanations of complex yielding behaviour in glassy states of
systems of particles with short-range attraction
\citep{pham3,PhamJRheol} can both be directly tested using such
methods. In other cases, the use of imaging may throw light on
well-known and long-studied phenomena that remain incompletely
understood, e.g. shear thickening of concentrated colloidal
suspensions \citep{fall}. The detailed comparative study of
distinct but related systems using imaging techniques, e.g.
Brownian and non-Brownian suspensions, or particulate suspensions
and emulsions, should also be a fruitful area for exploration.

Further technical developments will enhance the power of the
methodology. Faster scanning methods will obviously extend the
upper limit of flow rates that can be studied. The development of
algorithms for identifying \citep{penfold} and tracking particles
in polydisperse systems will significantly extend the range of
systems that can be studied -- there has been no quantitative
study to date of polydisperse suspensions under flow by imaging.
Further use of confocal imaging combined with (simultaneous) bulk
rheometry will give a wealth of information on the relation
between micro-scale dynamics and bulk flow and provide a crucial
test ground for various theories. The coupling of
three-dimensional imaging with microrheological techniques
utilising optical trapping (`laser tweezers' \citep{waigh}) should
also yield many new discoveries and insights; confocal microscopy
in two dimensions has already been used to explore the structure
of the `wake' downstream from a probe particle being dragged
through a concentrated suspension \citep{furst}. Studying flow in
microfluidic geometries using time-resolved single-particle
imaging is of direct relevance to many emergent lab-on-a-chip type
applications. And the parallel use of optical microscopy and other
imaging techniques (perhaps especially NMR) to study the same flow
system should open up many possibilities.

Finally, the example applications reviewed here pertain mostly to
well-characterised, model particles. Confocal microscopy
can fruitfully be used to image the (static) microstructure of
many `real-life' systems, such as foods \citep{foodconfocal}. To
date, there have been few confocal microscopy and rheological
studies of such systems \citep{nicolas}. Given the importance of
rheological properties in the processing and utilisation of a wide
range of industrial products (controlling the `mouth feel' of food
is but one example), we may expect increasing application of this
methodology to applied colloidal systems.

\section{Acknowledgements}

\label{sec:thanks}

RB, ABS and WCKP were funded respectively by EPSRC grants
GR/S10377/01, EP/E030173/1 and EP/D067650. LI acknowledges support
from EU network MRTN-CT-2003-504712.


\end{document}